\documentclass[prc,twocolumn,showpacs,showkeys,aps,secnumroman,amssymb,nobibnotes,balancelastpage,nofootinbib]{revtex4-1}

\usepackage{graphicx}
\usepackage{graphics}
\usepackage{dcolumn}
\usepackage{bm}
\usepackage{amsmath}
\usepackage{epsfig}
\usepackage{longtable}
\usepackage{xcolor}
\usepackage[normalem]{ulem}
\usepackage{multirow}
\usepackage{sidecap}
\usepackage [latin1]{inputenc}
\usepackage{color}

\newlength\figwidth\figwidth=0.5\textwidth
\renewcommand{\thefootnote}{\fnsymbol{footnote}}

\begin{document}

\title{High-resolution study of excited states in $^{158}$Gd with 
  (p,t) reactions.}

\author {A.~I.~Levon$^{1}$, D.~Bucurescu$^{3}$, C.~Costache$^{3}$, T.~Faestermann$^{2}$,
R.~Hertenberger$^{2}$, A.~Ionescu$^{3,4}$, R.~Lica$^3$, A.~G.~Magner$^{1}$,
C.~Mihai$^{3}$, R.~Mihai$^{3}$, C.~R.~Nita$^{3}$, S.~Pascu$^{3}$, K.~P.~Shevchenko$^{1}$, A.~A.~Shevchuk$^{1}$, A.~Turturica$^{3}$,  and H.-F.~Wirth$^{2}$
}

\affiliation{$^1$ Institute for Nuclear Research, Academy of Science, Kiev, Ukraine}
\email[Electronic address: ] {alevon38@kinr.kiev.ua}

\affiliation{$^2$ Fakult\"at f\"ur Physik, Ludwig-Maximilians-Universit\"at M\"unchen, Garching, Germany}

\affiliation{$^3$ H.~Hulubei National Institute of Physics and Nuclear
Engineering, Bucharest, Romania}

\affiliation{$^4$ University of Bucharest,
Faculty of Physics,  Bucharest-Magurele, Romania}

\begin{abstract}

  The excitation spectra in the deformed nucleus $^{158}$Gd have been studied
  with high energy resolution
by means of the (p,t)
 reaction using the
Q3D spectrograph facility at the Munich Tandem accelerator.
The angular distributions of tritons were measured for more than
 200 excited states seen in the triton spectra up to 4.3 MeV.
 A number of 36 excited 0$^+$ states (five tentative),
 have been assigned
by comparison of experimental angular distributions with the calculated ones using the CHUCK code.
Assignments for levels with higher spins are the following:   95 for 2$^+$ states, 64 for 4$^+$ states,
14 for 6$^+$ states and about 20 for negative parity states.
Sequences of states which can be treated as rotational
bands are selected. The analysis of the moments of inertia defined for these bands is carried out.
This  high number  of  excited states
in a deformed nucleus, close to a complete level scheme, constitutes a very
good ground
to check models of nuclear structure.   The large ensembles of states with the
same spin-parity offer unique opportunities
for statistical analysis. Such an analysis
for the 0$^+$, 2$^+$ and 4$^+$ states sequences, for
 all $K$-values and for well-determined projections $K$ of the angular
 momentum is performed. The obtained data may indicate on a K symmetry breaking.
 Experimental data are compared with interacting
boson model (IBM) calculations using the $spdf$
 version of the model. The energies
of the low-lying levels, the transition probabilities in the
first bands and the distribution in transfer intensity of
the 0$^+$ states are calculated and compared with experiment.
\end{abstract}
\bigskip
\date{\today}


\maketitle

\section{Introduction}
The nucleus $^{158}$Gd is located in a region of strong
deformation.
 Excitation
spectra of the even-even
nuclei in this region are complex. Collective excitations - both  of
the rotational and vibrational nature - are dominant.
The particle-hole nucleon excitations can also contribute
to such spectra.
 Interactions of all these sources of
 nuclear excitation
complicate the understanding of the resulting structures,
and therefore a full description has not been achieved yet.
In fact, nuclear collective  excitations even at low
energies still represent a challenge for the theoretical
models. At low excitations these states can be analyzed
in terms of the
beta vibrations, pairing vibrations, spin-quadrupole
interaction, shape coexistence, one- and two-phonon states, etc.
At higher
 excitations, one expects multi-phonon states
and mixing of all these excitations by the residual interaction.
 Detailed experimental data on the properties
 of many excited states of deformed nuclei over an extended
 excitation energy range are required in order to unravel these aspects.

\begin{figure*}
\includegraphics[width=0.92\textwidth,clip]{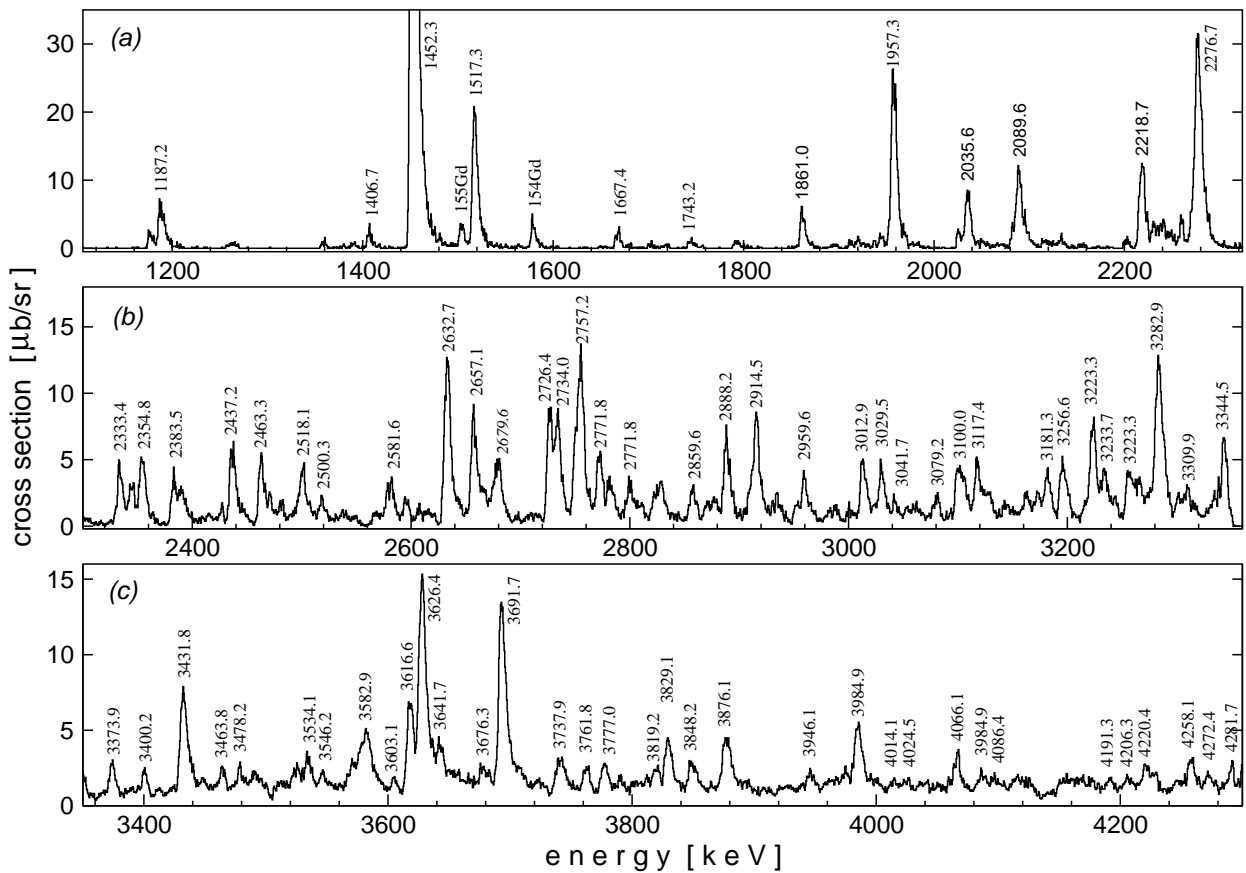}
\caption{{\small The triton spectrum from the $^{160}$Gd(p,t)$^{158}$Gd
reaction measured at angle 5$^\circ$.   Peak labels represent excitation energies in keV.}}
\label{Spec_1-4}
\end{figure*}

 Most detailed studies of
 the collective modes in the nucleus $^{158}$Gd were
performed in the radiative capture  \cite{Gre78,Bor99}
and in the (n,n$^\prime \gamma$) reaction \cite{Gov01}.
These studies were very important for a
complete determination of the level scheme at low spins  and
up to low-to-moderate level density,
 that corresponds to about
 2.5 MeV excitation.  Nearly 90 levels with low spins of
positive and negative parity up to 3 MeV
were identified in this region and many
of these states were
 combined into rotational bands. A total of
 thirteen excited rotational bands with band-head energies
below 1.8 MeV were incorporated in the level scheme.
They include the  octupole-vibrational bands with
band-heads 0$^-$ and 1$^-$, the $\gamma$-vibrational band and
three excited  $0^+$ bands. Several two-quasiparticle bands with
band-heads 4$^+$, 4$^-$ and 1$^+$ were identified too.
The study of $\beta^-$ decay of $^{158}$Eu \cite{Klu74} is most
informative among other radioactive decay studies, and has provided
31 excited states and 94 $\gamma$ transitions, all
incorporated in a level scheme. The coincidence measurements
have provided reliable branching ratios for members of
the $\gamma$-vibrational band and members of $K^\pi=0^-$
and 1$^-$ octupole bands. Precise excitation energies,
reduced transition probabilities and decay branching ratios
of numerous $I=1$ states were extracted from the energies and
angular distributions of the scattered photons in the nuclear
resonance fluorescence experiment \cite{Pit89}.
 The ground 0$^+$ band and octupole 1$^-$ band were extended
 to the 12$^+$ and 9$^-$ states, respectively, by Coulomb excitation
 \cite{Sug93}.
However, all these studies  had many difficulties  at states above
$\sim$2 MeV of excitation energy, and completeness of data was rapidly lost.

The most productive mode of
obtaining information about collective
and other excitation modes is the use of the direct reaction
of two-neutron transfer, which, for practical reasons, is
mainly the (p,t) reaction. It was found to be a very effective tool to study
the multiple 0$^+$ excitations in  actinide and rare earth nuclei
\cite{Ack94,Bal96,Lev94,Lesh02,Wir04,Lev09,Lev13,Lev15,Spi13,Mey06,Buc06, Bet09,Ili10}.
For some nuclei in these studies, extensive information was also obtained
for states with higher spins of  the positive and negative parity
up to 6  \cite{Buc06,Lev09,Lev13,Lev15,Spi18}.
So far, almost all the studies  with the (p,t) reaction
were performed for excitation energy below 3 MeV.
 The study
of 0$^+$ states up to about 4.2 MeV for $^{158}$Gd was
recently performed  in Ref.~\cite{Lev19},
and of 0$^+$ and 2$^+$ states in the case of $^{168}$Er (see Ref.~\cite{Buc06}).

Several theoretical approaches were aimed to explain the results
 obtained by these studies, e.g.,
the interacting boson model (IBM) \cite{Iach87,Casten01}
and its expansion  using the $s,p,d,f$ bosons \cite{Zam01,Zam03},
 the projected shell model (PSM) \cite{Har95,Sun03},
  the  quasiparticle-phonon model (QPM) \cite{Sol89,QPM,Lo04,Lo05},
  and a model including  the monopole
  pairing, the quadrupole-quadrupole 
  and spin-quadrupole forces in
the framework of the random phase approximation (RPA) \cite{Mur10}.
Both QPM and IBM predict a number of 0$^+$ states and a cumulative
cross section for their excitation which basically agreed with
experiment for low energies.
However,  both models fail to give a detailed explanation
of the individual states. Most excitations calculated in the IBM
have two $pf$ bosons in their structure, therefore being related
to the presence of a double octupole structure. At the same
time the QPM predicts only minor double-octupole phonon
components in states below 3 MeV.

This paper presents results of new measurements, with the $^{160}$Gd(p,t)$^{158}$Gd reaction,
of positive and negative parity states
in the region from 1.7 MeV up to 4.3 MeV excitation.
We identified in $^{158}$Gd  230 states with different spins in this
energy interval.
 The angular distributions of tritons were measured for 205 states
 seen in the triton spectra. Firm assignments of spins and parities
 have been obtained for  most of these excited states
by comparison of experimental angular distributions with
the calculated ones using the distorted wave Born approximation (DWBA).
 Sequences of states were selected that can be treated as
 rotational bands. They are used for statistical analysis
 of sequences of 2$^+$ and 4$ ^+$states  with different fixed $K$ projection
 of the angular momentum on the symmetry axes.
 A new approach is used for fitting the  nearest neighbor-spacing distributions (NNSD)
to investigate the fluctuation properties of the experimental spectra.
 The nature of 0$^+$ and other states is analysed
 in the frame of the IBM.

\section{\label{sec:ExAnRe} Experiment, analysis and   results}

\subsection{\label{sec:exp_det} Experimental details}
The experiments have  been performed at the Tandem accelerator of
the Maier-Leibnitz-Laboratory of the Ludwig-Maximilians-University
and Technical University of Munich using a 22 MeV proton beam. The
reaction products were analyzed with the high-precision Q3D
spectrograph. A long (1.4 m) focal-plane detector provides the
$\Delta E$/$E$ particle identification of the light  ejectiles
and position determination \cite{Wir01}. The different
runs were normalized to the beam current integrated into a Faraday
cup placed behind the target.

The  experiment in the high-energy region 3.0 - 4.3 MeV has been performed
on a 110 $\mu$g/cm$^2$ target of isotopically enriched $^{160}$Gd (98.10\%)
with a 14 $\mu$g/cm$^2$ carbon backing. Known impurities in the target material
consist of $^{158}$Gd (0.99\%), $^{156}$Gd (0.33\%), and $^{157}$Gd (0.44\%).
The resulting triton spectra have a resolution of 4 - 7 keV (FWHM) and
are background-free.
The  acceptance of the spectrograph
$\Delta \Omega$ was 14.43  msr  for all angles, except for
the most forward angle 5$^\circ$, where it was
7.50  msr.
Typical beam current was around 1.0 $\mu$A.
The angular distributions of the cross sections were obtained from
the triton spectra at eight laboratory angles from 5$^\circ$ to 40$^\circ$ in step of 5$^\circ$.
The low energy spectra in the interval from 0 to 3.4 MeV
have been  also measured at the angle of 5$^\circ$  for three magnetic setting,
which are all overlapping with the neighboring regions.
For the calibration of the energy scale, the triton spectra
  from  the reaction  $^{154}$Gd(p,t)$^{152}$Gd  have been measured at
the same magnetic  setting.  In this way, the  high energy spectrum of $^{158}$Gd
was calibrated  by the known energies  of the nucleus $^{152}$Gd.

The  experiment  in the low-energy region 1.7 - 3.2 MeV was performed
with a 125 $\mu$g/cm$^2$ target of $^{160}$Gd.
The acceptance 
$\Delta \Omega$
was 9.8 msr  for 6$^\circ$
and 14.5 msr for other angles.
The resulting triton spectra have a slightly lower resolution
of 8 - 9  keV (FWHM).
For the calibration of the energy scale, the triton spectra from
the reaction $^{172}$Yb(p,t)$^{170}$Yb were measured at the same magnetic settings.
The low-energy spectrum calibrated in such a way has
a 250 keV overlap with the high-energy spectrum fixed by the previous experiment. Many levels
of $^{158}$Gd well-known from the resonance capture and from
the (n,n$^\prime \gamma$) reaction
are correctly fitted with this calibration in  the low energy region.
The spectra in low and high energy
intervals calibrated by the corresponding reactions $^{154}$Gd(p,t)$^{152}$Gd
and $^{172}$Yb(p,t)$^{170}$Yb coincide in the overlapping region.
The difference in the energies determined by these  calibrations
in the overlapping region does not exceed 1 keV.

The details of the experiment and especially those of
 the energy calibration procedure are given in Ref.~\cite{Lev19}
  which deals with the study of excited 0$^+$ states in $^{158}$Gd.
Some results of the (p,t) experiment at  low energies
performed by a Yale-Munich-K{\"o}ln-Bucharest collaboration
(the YMKB experiment) \cite{Mey06}  were
also analysed in this publication.

 Fig.~\ref{Spec_1-4}(a-c) shows the triton spectrum  over the
  energy interval from 1.0 to 4.3 MeV, taken  at the detection angle of 5$^\circ$.
Some strong peaks are labeled by their energies in keV.

The analysis of triton spectra was performed  by using
the program GASPAN \cite{Rie91}.
Peaks of the spectra which are measured at 5$^\circ$ degree
have been  identified for 230 levels,
though the angular distributions  for  all eight angles could be measured
only for  205 levels.
The differential cross sections were calculated
 by the following equation

\begin{eqnarray}
\frac{d\sigma(\theta)}{d\Omega} = \frac{N(\theta)}{\Delta\Omega \times I_{total}
 \times D_{target} / \cos(\theta)}
\end{eqnarray}

Here $N(\theta)$ is the number of tritons measured for each state at a Q3D angle $\theta$,
corrected for the dead time of the data-acquisition system,
$\Delta \Omega$ is the acceptance of the spectrograph, $I_{total}$ is the total
number of protons measured by the Faraday cup, and $D_{target}$/cos($\theta$)
is the effective target thickness. The angle $\theta$ is also the angle
between the target area and the beam  axis.
To determine the integrated (p,t) excitation cross
section, the differential cross sections were integrated over the
covered angular range.

\subsection{\label{ssec:DWBA} DWBA analysis }

To determine the value of the transferred angular momentum $L$ and
spin ($I = L$) for each level in the final nucleus $^{158}$Gd, the
observed angular distributions are compared with calculations using
the DWBA. The coupled-channel
approximation (CHUCK3 code of Kunz \cite{Kun}) and the optical
potential parameters suggested by Becchetti and Greenlees
\cite{Bec69} for protons and by Flynn et al. \cite{Fly69} for
tritons have been used in the calculations.

 In principle, the
transfer of the two neutrons coupled to spin 0 should contain the
contribution of different $j$
spins of the two particles.
The orbitals close to the Fermi surface have been used as the transfer
configurations.  For $^{158}$Gd and $^{160}$Gd,  such configurations
include the orbitals
which correspond to those
 in the spherical potential, namely,
  $2f_{5/2}$, $1h_{9/2}$, $1h_{11/2}$, and $1i_{13/2}$.
  Since we do not know the dominant
transfer for each state, all of them were tested to get a better fit
of the experimental angular distributions.
 The angular distributions for
the 0$^+$ states are reproduced very well by  a one-step process.
Only two configurations in possible combinations have been taken into
account,
 that simplifies the calculations. The experimental results
and the details of the DWBA calculations for 0$^+$ states are
presented in the publication
\cite{Lev19}. Thirty-two new excited 0$^+$ states
(four tentative) have been
assigned up to the 4.3 MeV excitation energy.  Thus,
the total number of
0$^+$ excited states, besides the ground state (g.s.) in
$^{158}$Gd, was increased
up to 36, the highest number of such states observed so far
in a single nucleus.

\begin{figure}
\includegraphics[width=0.48\textwidth,clip]{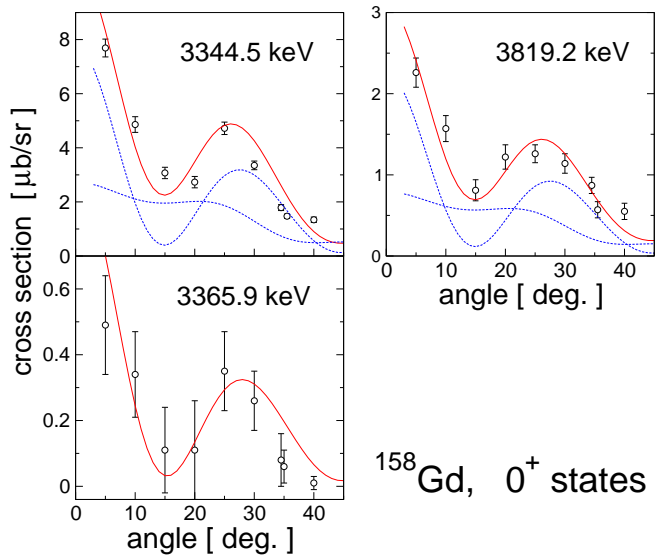}
\vspace{-15.0mm}
\caption{{\small The 0$^+$ state at 3365.9 keV additionally identified in this study
and suggested fits for the states 3344.5 and 3819.2 keV.  The blue lines represent
the result of calculations for the 0$^+$ and 4$^+$ states, respectively,
the sum of which
fits the experimental angular distributions. }}
\label{fig_0+_add}
\end{figure}

In the  present detailed
analysis,  an
additional weak 0$^+$ excitation at 3365.9 keV was identified.
The angular distribution for this state is shown in Fig.~\ref{fig_0+_add}.
Another problem met in the previous study \cite{Lev19}
is  a tentative 0$^+$ assignation for two states at 3344.5 and 3819.2 keV.
For these states the reason of this tentative assignment is
the absence of a deep minimum at an angle of about 17$^\circ$
(Fig.~\ref{fig_0+_add}).
The calculated angular distribution has such
a form at the transfer of a pair of $i_{13/2}$ neutrons but only for
a lower excitation energy. It proved impossible
to fit well the experimental angular distributions by
using the actual reaction energies in such calculations.
Calculations for transferring other
angular momenta do not allow to describe the experimental
angular distributions, and thus, rule out other spin assignments.
There is another possible explanation for this shape of the angular distribution:
 the overlap with another level having a very close energy.
The overlap of the angular distributions for the  0$^+$ state with
those for a 4$^+$ state  explains the experimental angular distributions
 for both levels as demonstrated in Fig.~\ref{fig_0+_add}.
 Of course, this is only a tentative explanation.

\begin{figure}
\includegraphics[width=0.45\textwidth,clip]{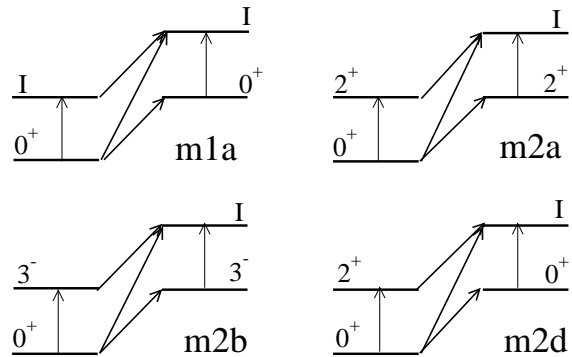}
 \caption{{\small Schemes of the CHUCK3 multi-step
    calculations tested with spin assignments of  excited states
    in $^{158}$Gd (see Table~\ref{Tab:expEI}).}}
    \label{fig_schem}
\end{figure}


\begin{figure}\centering
\includegraphics[width=0.475\textwidth,clip]{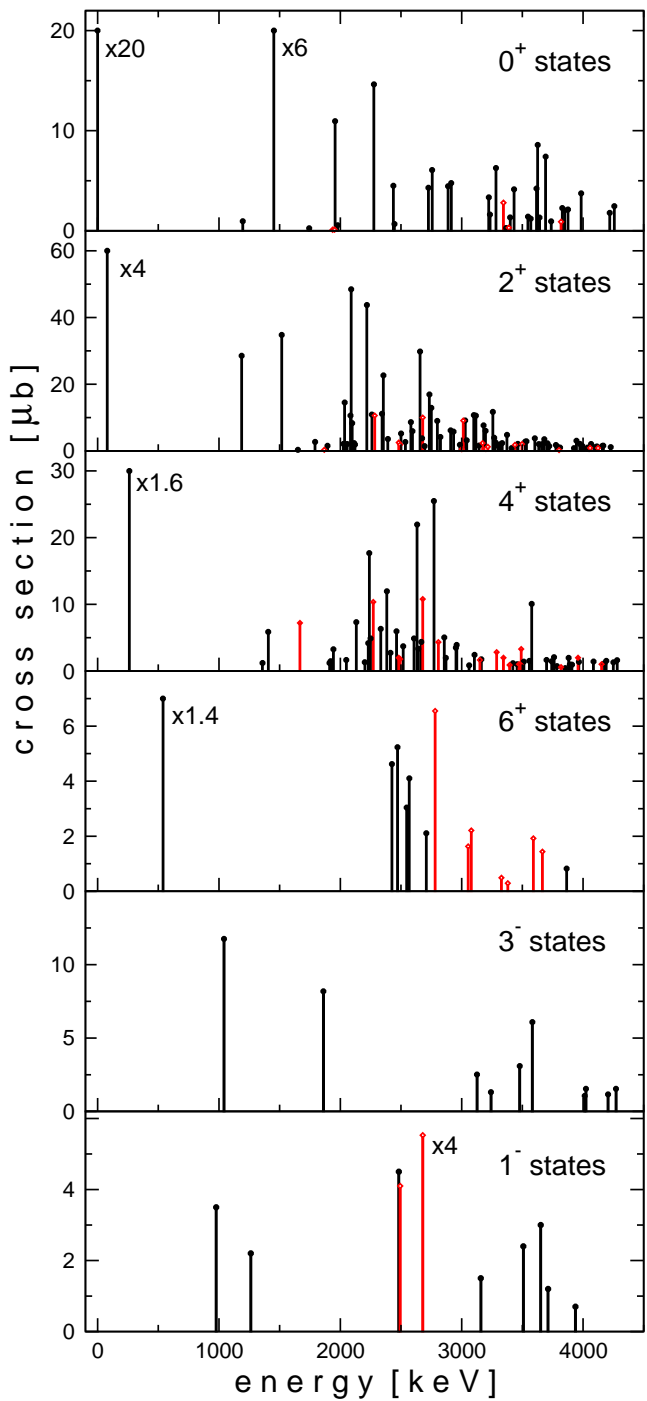}
 \caption{{\small The (p,t) strength integrated in
 the angle region 0$^\circ$ - 45$^\circ$ for 0$^+$, 2$^+$, 4$^+$, 6$^+$,
 3$^-$ and 1$^-$ states  in $^{158}$Gd.
 The levels identified reliably and tentatively are indicated by black
 filled circles and
 by red open diamonds, respectively.}}
    \label{Integr_Int}
\end{figure}

The situation is more complex for the states with higher spins.
Only a few experimental angular distributions could be fitted by
the calculated ones for the one-way direct transfer of two neutrons
with nonzero orbital angular momentum.
 The angular distribution for such states may be altered
 due to inelastic scattering (coupled channel effect),
 treated here as multi-step processes. Taking into account these
  circumstances, one can obtain spin assignments
  for  most excited states in the final nucleus $^{158}$Gd by fitting the angular
distributions obtained in the DWBA calculations to the
experimental ones.
The multi-step transfer schemes  used in the present DWBA
calculations are displayed in Fig.~\ref{fig_schem}. The best fit is 
achieved by changing the amplitudes of each branch in  the
multi-step transfer. The shape of the angular distribution in this
case may be drastically different from the shape of that for the
one-way transfer. Moreover, with the projectile
energy used in the experiment,
 the shape of the one-step angular distribution also
 changes   with increasing of the  excitation  energy (see below).
Nevertheless, the selectivity of such  spin
assignments is quite reliable. The spins assigned in such a way are
confirmed by comparison with the spin values well-defined in other
experiments.

\begin{figure}
\includegraphics[width=0.48\textwidth,clip]{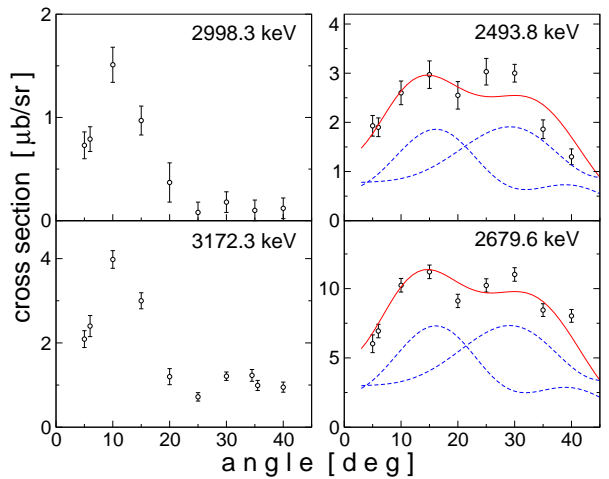}
 \caption{{\small Angular distributions for the states
 with problematical fits.
 See text for details.}}
    \label{Nonfit}
\end{figure}

The results of this study concerning all the states identified in
the (p,t) reaction are collected in Table~\ref{Tab:expEI}.
 They are also presented in a compressed form in Fig.~\ref{Integr_Int}.
For the states below 1743.2 keV we
obtained only the absolute cross sections at  5$^\circ$  because
the angular distributions  themselves were not measured.
Therefore, their spins were not assigned in this work
and are not shown in  Table~\ref{Tab:expEI}.
Excitations of the 0$^+$ and 2$^+$ states
in the nuclei of the impurity isotopes in the target material
manifest themselves in the observed triton spectrum.
The 1577 keV excitation is important in our study.
The 0$^+$ assignment at 1576.932(16) keV from the (n,$\gamma$)
reaction \cite{Gre96}  was confirmed
in the (p,t) reaction \cite{Lesh02}.  However,
later,  no $\gamma$-rays were detected as
decaying the level 1577 keV  when
studying the 0$^+$ states
in the (n,n$^\prime\gamma$) reaction \cite{Lesh07}.
The triton energy associated with this level is
near that for the g.s.
in the $^{156}$Gd(p,t)$^{154}$Gd reaction.
Therefore, the corresponding peak can be interpreted as
an excitation  on the $^{156}$Gd contamination in
the target material.
The observed cross section 5.6 $\mu$b/sr is  somewhat smaller
than the calculated 7.8 $\mu$b/sr when using the cross section
for the $^{156}$Gd(p,t)$^{154}$Gd reaction from Ref.~\cite{Mey06}.
Thus, the present (p,t) data do not  confirm presence
of the 1577 keV level in the nucleus $^{158}$Gd.

Spins and parities for ten
states above 1743 keV are not shown in  Table~\ref{Tab:expEI}.
The energies of these
 states were determined in the spectrum at 5$^\circ$
measured with good statistical accuracy.  However, identification
of the corresponding peaks in the spectra for other angles was difficult
and consequently their angular distributions could not be measured.
The shape of the angular distributions for two states,  2998.3 and 3172.3 keV,
could not be attributed to any calculated angular distribution
(Fig.~\ref{Nonfit}).
However, since  the beginning of the angular distributions is close
to that for the 2$^+$ and 1$^-$ states, these spins were assigned tentatively for these states.
Finally, the angular distributions for two states at 2493.8 and 2679.6 keV
can be fitted by calculated ones for one-way transfer to a 1$^-$ state.
However, their cross sections are excessively high as compared
with other 1$^-$ states observed in $^{158}$Gd.
Therefore, an alternative description
of the angular distribution can be considered.  Namely, the superposition of
two distributions, for 2$^+$ and 4$^+$ states, as shown in Fig.~\ref{Nonfit}.
That is, the corresponding peaks in the triton spectrum are assumed to be
doublets.  Both options are included in Table~\ref{Tab:expEI} as tentative assignments.

The ground state   rotational-band members are excited up
to $8^+$ in such experiments \cite{Lev09,Lev13,Lev15,Spi13,Spi18}
(for  $^{158}$Gd the  8$^+$ state peak
is overlapped by the peak of the excitation
of the g.s. $^{156}$Gd impurity).  Nevertheless, angular distributions
could be measured up to $6^+$. As one can see
from Fig.~\ref{Integr_Int}, the cross section is steadily
decreasing with increasing spin.
 Figs.~\ref{2plus}, \ref{4plus} and \ref{6plus} show the experimental data
 for the angular distributions for 2$^+$, 4$^+$, 6$^+$, as well as for
 1$^-$, 3$^-$ states, all given in $\mu$b/sr
 and their values are plotted with symbols with error bars while the
Q-corrected CHUCK3 calculations are shown
 by full lines. The solid (red) lines present the firm assignments
and the solid (blue) lines show  tentative assignments.
Fitting of the calculated angular distributions to the experimental ones
allowed  to determine the spins and parities for  most of
final states which were identified.

\subsection{\label{spec} Some specific features of 
  angular distributions
in the extended energy range.}

\textsl{0$^+$ states}.  Excitations
of 0$^+$ states
are  possible only in the one-way transfer of
a pair of neutrons. The
shape of angular distribution depends  only slightly on
the neutron configuration and is characterized by a steeply rising
cross section at small angles,
 a sharp minimum at angles of 10$^\circ$-17$^\circ$ and a weak maximum at angles
 of 25$^\circ$-35$^\circ$.  A
significant shape deviation for $^{158}$Gd was observed only
for two excitation energies and is tentatively explained by a possible
overlap with
 the angular distribution of another state (see above).

\begin{figure*}
\begin{center}
\epsfig{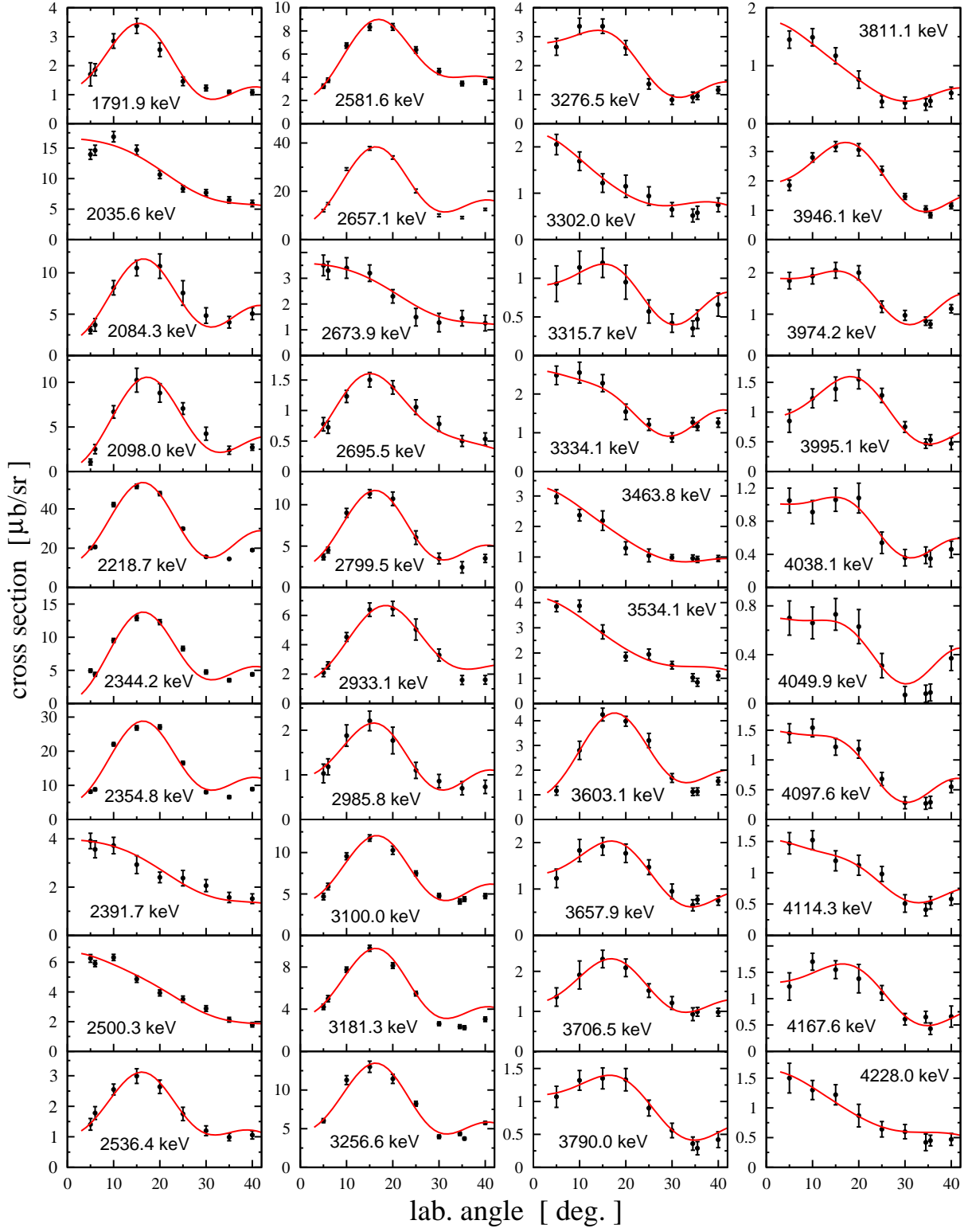}
    \caption{\label{2plus}   Angular distributions of assigned
    $2^+$ states in $^{158}$Gd and their fit with  CHUCK3  calculations.
     The  $(ij)$ transfer configurations and schemes used
    in the calculations for the best fit are given in Table~\ref{Tab:expEI}.}
\end{center}
\end{figure*}

\textsl{2$^+$ states}. The  angular distribution for the 2$^+$  states calculated
for  the one-way transfer of a pair of neutrons has a "bell-shaped" form with a deep minimum
at small angles and a maximum at angles of 15$^\circ$-18$^\circ$. As
 an example one can see the distribution for  the
energy of 2218.7 keV in Fig.~\ref{2plus}.
The experimental angular distributions have such a form for many excitation energies.
However, the detailed fitting  needs in some cases
at least small inclusions of two-step processes
 involving inelastic scattering through intermediate states. The calculated and experimental
 angular distributions also change the shape with increasing
 excitation energy, even for the one-way transfer.
 The cross section at small angles gradually increases  with
 increasing excitation energy up
 to the maximum values at angles of 15$^\circ$-18$^\circ$. This can be seen, for
 example, already for
 the energy of 3315.7 keV in Fig.~\ref{2plus}.
 A special case
 is  represented by excitations in which
 inelastic scattering through  intermediate states in the two-step processes
 plays a significant or even dominant role.
 As an example, such a case is the excitation of the 2$^+$ state in the
 ground state band.
 In this case, the angular distribution has a strong maximum at small angles, but, unlike the case of the 0$^+$ states, there is not a deep minimum.
 The 2$^+$ assignments in cases of such
 angular distributions are confirmed by known spins in
 previous studies \cite{Lev09,Lev13,Lev15}  for the states 2500.3
  and 2673.9 keV in this work (see Fig.~\ref{2plus}).

\begin{figure*}
\begin{center}
\epsfig{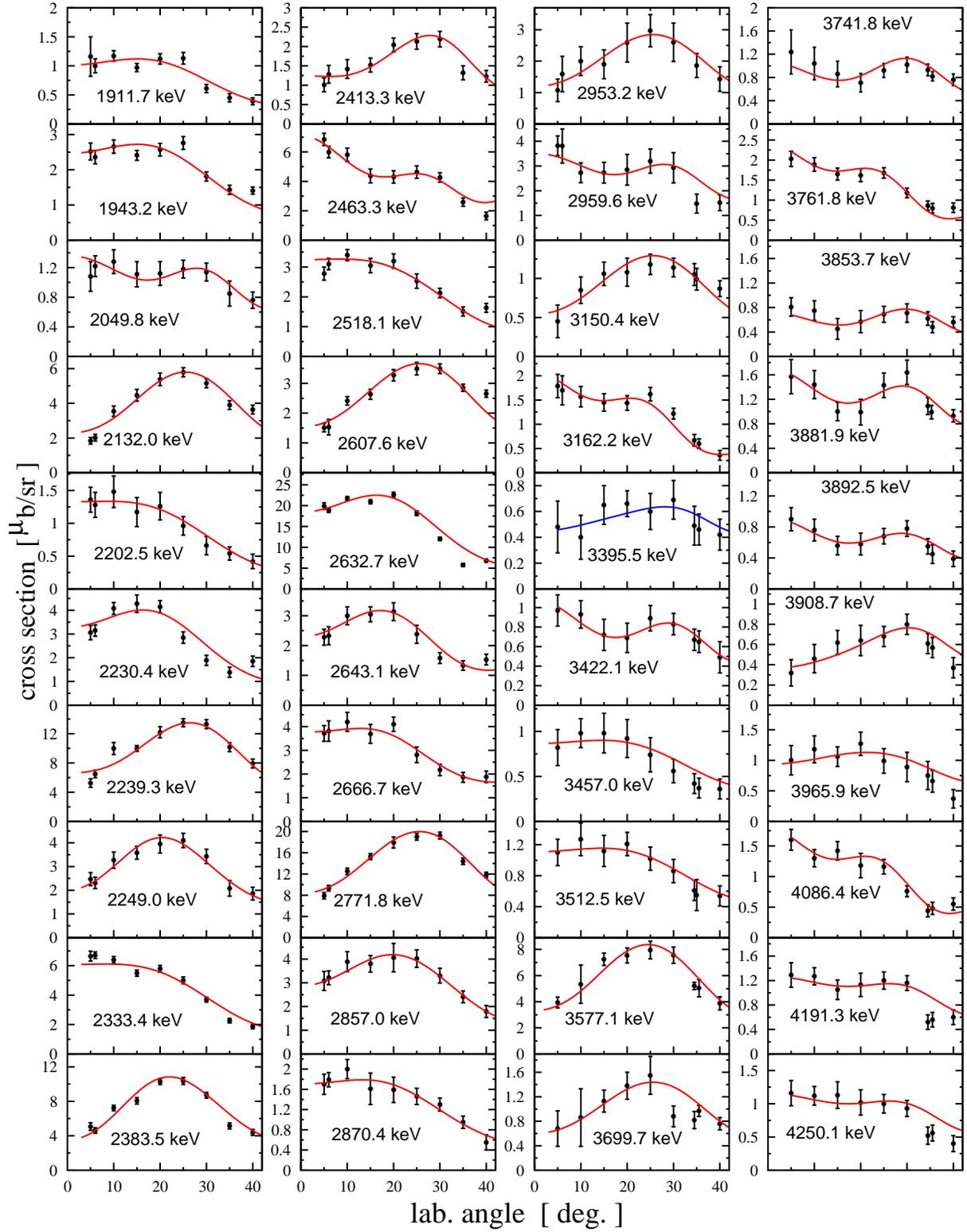}
    \caption{\label{4plus} Angular distributions of assigned
    $4^+$ states in $^{158}$Gd and their fit with  CHUCK3  calculations.
     The  $(ij)$ transfer configurations and schemes used
    in the calculations for the best fit are given in Table~\ref{Tab:expEI}.
     The red lines indicate firm assignments and the blue lines are putative ones.}
\end{center}
\end{figure*}

 \textsl{4$^+$ states}.  The angular distributions for the 4$^+$ states are reproduced
 with small admixture of two-step processes involving inelastic scattering of intermediate
 states only for some excitation energies, as for example for the state at 2049.8 keV in Fig.~\ref{4plus}.
 With increasing excitation energy,  the calculated and experimental
 angular distributions even at
 the one-way transfer change the shape similar to 2$^+$ states.
 The cross section at small
 angles gradually decreases  and there is an increasing maximum at an
 angle of about
 30$^\circ$.
 It is seen in Fig.~\ref{4plus} for instance
 for the excitation energy of 2132.0 keV.
 Similarly to the 2$^+$ states, a special case is represented
 by the
 excitation in which the multi-step processes play a significant role.
 In this case, the angular distribution has a maximum at small angles,
 although  it is  not as
 pronounced as that for 2$^+$ states, while
  the deep minimum of the 0$^+$ states is absent.
  It is seen, e.g., in Fig.~\ref{4plus},
  for the excitation energy 2202.5 keV.

\begin{figure*}
\begin{center}
\epsfig{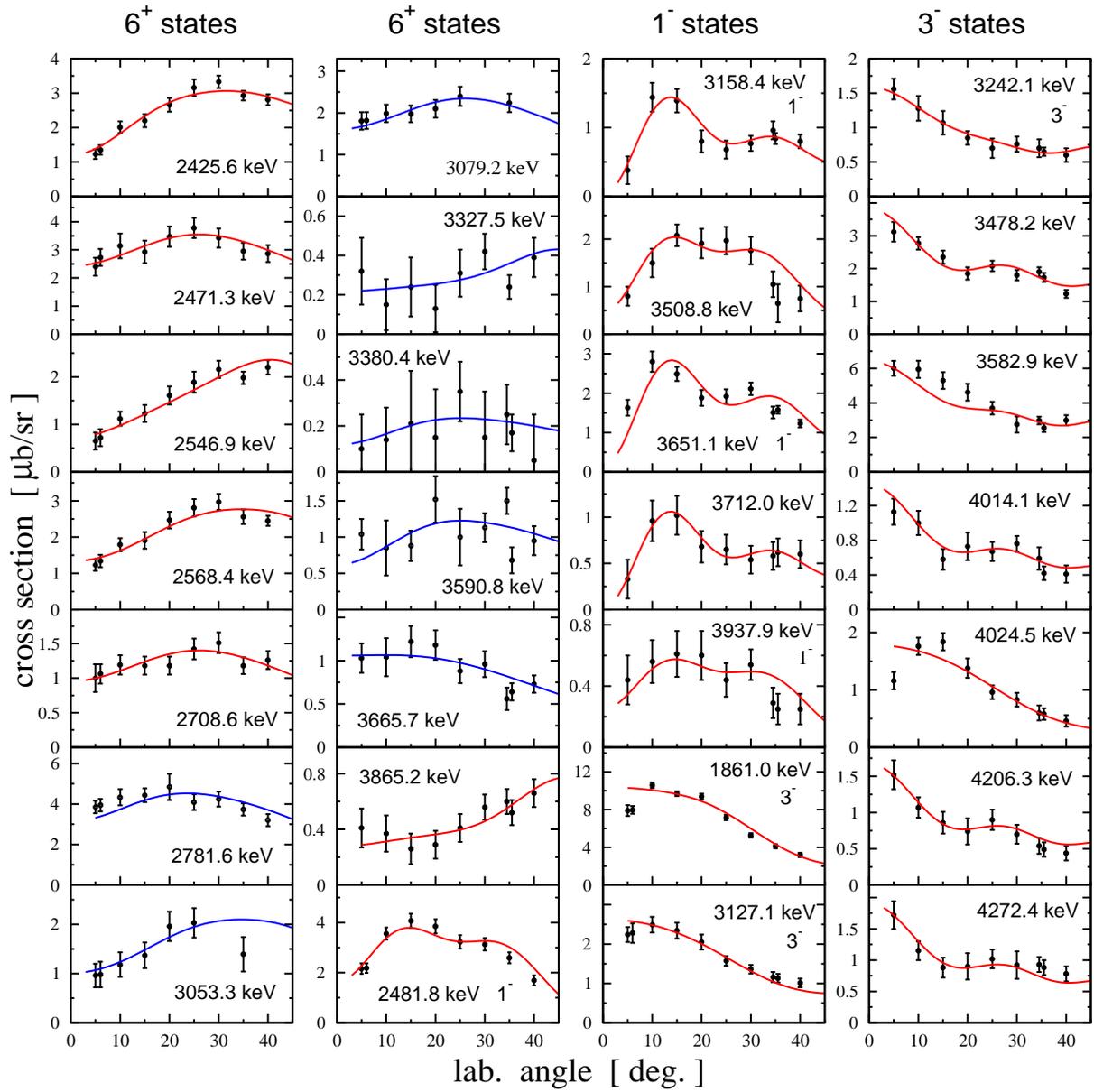}
    \caption{\label{6plus} Angular distributions of assigned
    $6^+$, $1^-$ and $3^-$ states in $^{158}$Gd and their fit with  CHUCK3  calculations.
     The  $(ij)$ transfer configurations and schemes used
    in the calculations for the best fit are given in Table~\ref{Tab:expEI}.
    The red lines indicate firm assignments and the blue lines are putative ones.
    The red dashed lines are calculations with changed potential parameters for tritons (see text).}
\end{center}
\end{figure*}

 \textsl{6$^+$ states}. The calculated angular distributions for 6$^+$ states with
 small admixture of two-step processes have  a pronounced maximum
 at the angle
 of about 45$^\circ$ at transfer of the f$_{5/2}$, h$_{9/2}$ and
 h$_{11/2}$ neutron pairs (the energy of 2546.9 keV in
 Fig.~\ref{6plus} as an example),
 and almost flat shape at transfer of i$_{13/2}$
 neutron pair  (energy of 3327.5 keV
 in Fig.~\ref{6plus} as another example). Taking
 into account  two-step processes leads
 to a shift of the maximum to smaller  angles.

 \textsl{1$^-$ states}. The angular distributions for the
 1$^-$ states are reproduced
 by the calculated ones for the  one-step transfer.
 They have two pronounced maxima and,
 therefore the assignment is reliable despite rather small cross sections of their excitation (Fig.~\ref{6plus}).

 \textsl{3$^-$ states}. The angular distributions for the
  3$^-$ states are reproduced
 by the calculated ones for
 the  one-step transfer for  most
 excitation energies
 (the energy of 3478.2 keV in Fig.~\ref{6plus} as  an example).
 Only some of them need a small inclusion of the two-step processes.
 The maximum of such  a distribution is found at  the angle of
 0$^\circ$ with the exception
 of two energies of 1861.0, 3127.1 and 4024.5 keV. The maximum of the angular
 distribution for these energies
  occurs at an angle of about 15$^\circ$, and  is not fitted by
 calculations with the potential parameters used for all other states.
 The  spin 3$^-$ of the first such state is well known from previous
 studies \cite{Nic17}.
 Therefore, this spin is assigned also for other two states.
  Minor changes of the
  parameters for tritons helped to fit these angular distributions,
namely the use of the triton potential parameters suggested by Becchetti and Greenlees \cite{Bec71}.

\newcolumntype{d}{D{.}{.}{3}}
\renewcommand{\thefootnote}{\fnsymbol{footnote}}
\begin{longtable*}{l c l c c c r}
\caption{\label{Tab:expEI} Energies of  levels in $^{158}$Gd,  spin
  assignments from the CHUCK3 analysis, the (p,t) reaction cross
  sections at 5$^\circ$, as well as integrated cross section over the measured
  values (i.e. 5$^\circ$ to 40$^\circ$), and the reference to the
  schemes used in the DWBA calculations. }\\
\hline\hline
\smallskip\\
\multicolumn{2}{c}{ENSDF Ref.~\cite{Nic17}}  \enspace & \multicolumn{4}{c}{Present data} \enspace & \enspace\enspace Way of\\
\smallskip\\
Energy [keV] & \enspace $I^\pi$   & \enspace\enspace Energy [keV]
 & \enspace $I^\pi$ & \enspace {d$\sigma$/d$\Omega$} & \enspace {$\sigma_{\mbox{integ.}}$}[{$\mu$b}]& \enspace\enspace fitting\\
&&&& at 5$^\circ$ [{$\mu$b/sr}] &&\\
\hline
\smallskip
\endfirsthead
\caption{Continuation}\\
\hline\hline
\smallskip\\
\multicolumn{2}{c}{ENSDF Ref.~\cite{Nic17}}  \enspace &  \multicolumn{4}{c}{Present data} \enspace & \enspace\enspace Way of\\
\smallskip\\
Energy [keV]  & \enspace $I^\pi$  & \enspace\enspace Energy [keV]
 & \enspace $I^\pi$ & \enspace {d$\sigma$/d$\Omega$} & \enspace {$\sigma_{\mbox{integ.}}$}[{$\mu$b}]& \enspace\enspace fitting\\
&&&& at 5$^\circ$ [{$\mu$b/sr}] &&\\
\hline
\smallskip
\endhead
\hline
\endfoot
\endlastfoot
~~~~~0.00		& 0$^+$	 &  ~~~~~0.1\it3	&&	1435 \it12 && \\						
~~~79.514	    & 2$^+$	 &  ~~~79.3 \it3	&&	267 \it4 && \\
~~261.458		& 4$^+$	 &  ~~260.1	\it3    &&  51.2 \it2 && \\
~~539.022		& 6$^+$	 &  ~~538.8 \it5    &&	1.9 \it3 && \\
$^{156}$Gd   	& g.s.	 &  ~~904.2	\it3	&&	11.7 \it7 && \\
~~977.156 \it2	& 1$^-$	 &  ~~977.3 \it4	&&	2.0  \it4	&& \\
$^{156}$Gd      &  2$^+$ &  ~~992.9	\it12	&&	2.1  \it4	&& \\
1023.698 \it3	& 2$^-$	 & 1023.4 \it12		&&  0.2  \it2 && \\
1041.640 \it3	& 3$^-$	 & 1041.6 \it3		&& 12.8  \it7 && \\
1176.481 \it5   & 5$^-$	 & 1176.7 \it5    	&&	3.5  \it4 && \\
1187.148 \it3	& 2$^+$	 & 1187.4 \it4		&& 11.4  \it7 && \\
1196.164 \it7   & 0$^+$	 & 1196.1 \it8 		&&  3.3  \it4 && \\
1259.870 \it2 	& 2$^+$  & 1260.8 \it8      &&  0.6 \it3  && \\
1263.515 \it3   & 1$^-$	 & 1262.7 \it6      &&	1.1  \it3 && \\
1358.472 \it3	& 4$^+$	 & 1358.4 \it4		&&  1.3  \it3 && \\
1380.634 \it6   & 4$^+$  & 1379.7 \it12     &&  0.4 \it3  && \\
1406.702 \it3   & 4$^+$     & 1406.4 \it3	&&  4.0  \it5 && \\
1452.353 \it6   & 0$^+$	 & 1452.3 \it6      &&	423   \it6 && \\
$^{155}$Gd      & g.s.      & 1503.3 \it3	&&  6.2  \it6 && \\
1517.480 \it3   & 2$^+$	 & 1517.3 \it10		&& 37.9  \it14 && \\
 $^{155}$Gd     & 5/2$^-$   & 1563.5 \it20  &&  0.4  \it3 && \\
$^{154}$Gd     	& g.s.	 & 1577.0 \it4      &&  5.6  \it6 && \\
1576.932 \it16	& 0$^+$	 &                  &&  && \\
1653			& 2$^+$     & 1650.0 \it24	&&  0.4  \it3 && \\
1667.373 \it6   & (4$^+$)   &	1667.3 \it4 &&  4.0  \it5 && \\
$^{154}$Gd      & 2$^+$     & 1701.4 \it12	&&  1.0  \it3 && \\
1716.807 \it5   & 5$^-$	 & 1717.9 \it15     &&  0.7  \it3 && \\

1743.147 \it14  & 0$^+$  & 1743.2 \it2   & 0$^+$  & 1.9 \it2 & 0.8 \it2 & sw.h09\\
1791.797 \it9   & 2$^+$  & 1791.9 \it5	 & 2$^+$  & 1.9 \it4 & 2.9 \it3 & sw.ii\\
1861.281 \it7   & 3$^-$	 & 1861.0 \it4	 & 3$^-$  & 8.9 \it7 & 10.0 \it4 & m2a.h09	\\
			    &        & 1868.1 \it8	 & (2$^+$)& 0.6 \it4 & 0.9 \it4 &  m1a.h09 \\
1894.578 \it21	& (2$^+$) &	1894.4 \it8  & 2$^+$  & 0.9 \it3 & 1.7 \it4 & m1a.h09	\\
                &        & 1911.7 \it8	 & 4$^+$  & 1.3 \it4 & 1.3 \it4 & m1a.ii \\
1920.264 \it6	& 4$^+$	 & 1920.9 \it6	 & 4$^+$  & 2.0 \it4 & 1.6 \it2 &	m1a.h11 \\
1935.5	 \it6	& 0$^+$	 & 1936.5 \it15	 & (0$^+$) & 1.0 \it2 & 0.3 \it1 & sw.h09i\\
                &        & 1943.2 \it8	 & 4$^+$  & 2.7 \it5 & 3.6 \it3 & m1a.ii \\
1952.425 \it25  & (0$^+$) &	1952.2 \it1	 &	      & 0.4 \it5 &           &  \\
1957.27	 \it9	& 0$^+$	 & 1957.3 \it3	 & 0$^+$  & 39.0 \it10	& 11.0 \it8 & sw.fi\\
1964.12 \it2    & 2$^+$  &               &         &            &            & \\
1972.2	 \it31  & (0$^+$) & 1977.6 \it12 & 0$^+$  & 1.3 \it2 & 0.6 \it2 & sw.h11i\\
                &        & 2026.3 \it8	 & 2$^+$  &	3.5 \it5 & 2.3 \it4 & m1a.h11 \\
2035.70 \it3    & 2$^+$	 & 2035.6 \it5   & 2$^+$&	15.2 \it9& 14.5 \it10& m1a.ii \\
                & 	     & 2049.8 \it10	 & 4$^+$  &	1.2 \it3 & 1.7 \it3 & m2a.h11 \\
                &        & 2056.5 \it8   & 2$^+$  &	1.2 \it4 & 2.2 \it4 & sw.h09 \\
2083.639 \it24	& 2$^+$	 & 2084.3 \it6   & 2$^+$  & 3.3 \it5 & 10.6 \it16 & sw.h09 \\
2089.254 \it8   & 2$^+$	 & 2089.6 \it5	 & 2$^+$  & 15.3 \it8 & 52.7 \it24	& m1a.h09 \\
2095.20  \it16  & (4$^+$)& 	 & &	& & \\
                &        & 2098.0 \it1	 & 2$^+$  &	1.1 \it3 & 8.3 \it12 & sw.h09 \\
                &        & 2113.5 \it6   & 2$^+$  &	1.0 \it2 & 2.5 \it5 & sw.h09 \\
2120.25 \it4    &        & 2120.8 \it8   & 2$^+$ &	1.0 \it2 & 2.1 \it3 & m1a.h09 \\
2134    \it7	&        & 2132.0 \it6   & 4$^+$  &	2.0 \it2 & 7.3 \it5 & m1a.h11 \\
2153.178 \it9   & (2,3)$^+$	& 2153.4 \it10& 3$^+$ & 0.6 \it1 & 3.0 \it4 & m2a.h09 \\
                &        & 2202.5 \it5   & 4$^+$  &	1.5 \it2 & 1.4 \it3 & m1a.h11 \\
                &        & 2218.7 \it5   & 2$^+$  &21.8 \it6 & 47.6 \it10& sw.h09 \\
                &        & 2230.4 \it6   & 4$^+$  &	3.6 \it3 & 4.2 \it4 & m1a.ii \\		
                &        & 2239.3 \it5   & 4$^+$  &	5.7 \it4 & 17.7 \it10& m1a.h11 \\
2249.61 \it5    & 2$^+$,3,4$^+$	& 2249.0 \it6 & 4$^+$ & 2.7 \it3 & 5.3 \it5 & m1a.h11 \\
2260.162 \it18 & 1,2$^+$ & 2260.3 \it5   & 2$^+$  &	5.2 \it3 & 11.9 \it7 & sw.h09 \\
2269.269 \it14  &(0,1,2)$^+$ & 2271.8 \it10& (4$^+$) & 8.3 \it3 & 11.3 \it16 & m1a.h11  \\			
2276.76 \it5    & 0$^+$   & 2276.7 \it4	 &  0$^+$    &52.3 \it15 & 14.6 \it9	& sw.h09\\			
2283.2  \it6	&	     & 2283.4 \it10  &  (2$^+$)  &10.2 \it12 & 11.5 \it8 & sw.ih\\
                &        & 2333.4 \it5	 & 4$^+$  &	7.2 \it4 & 6.9 \it3 & m1a.ii \\		
2340.3  \it3    & 2$^+$  &               &        &            &           &	\\										
2344.7  \it5	& 2$^+$,3$^+$ &	2344.2 \it5& 2$^+$&	5.4 \it3 & 12.1 \it5 & m1a.h09 \\		
2355.0	\it5    & 1$^+$,2$^+$ &	2354.8 \it4& 2$^+$&	9.1 \it4 & 22.6 \it8& m1a.h09 \\		
2384            &        & 2383.5 \it4  & 4$^+$	  & 5.5 \it3 & 13.0 \it6 & m1a.h11 \\
                &        & 2391.7 \it5	& 2$^+$   &	4.3 \it3 & 3.6 \it4 & m1a.ii \\		
                &        & 2413.3 \it8  & 4$^+$   &	1.1 \it2 & 2.7 \it3 & m1a.h11 \\		
                &        & 2425.6 \it8	& 6$^+$   &	1.3 \it2 & 5.0 \it3 & m2b.i13 \\		
                &        & 2437.2 \it4  & 0$^+$   &	11.9 \it4& 4.5 \it3 & sw.h09\\
2446.49 \it15	& 1	     & 2445.9 \it8	& 0$^+$	  & 1.5 \it2 & 0.7 \it3 & sw.h09 \\		
                &        & 2463.3 \it5  & 4$^+$   &	7.4 \it4 & 6.5 \it6 & m1a.ii \\		
                &        & 2471.3 \it6	& 6$^+$   &	2.6 \it3 & 5.7 \it6 & m2a.h09 \\		
2480.5 \it14	&        & 2481.8 \it6	& 1$^-$   & 2.4 \it2 & 4.8 \it5 & sw.hi \\			
                &        & 2493.8 \it10 & (1$^-$) & 1.4 \it2 & 4.1 \it4 & sw.h09i\\
                &        &          or  & ($2^+$+4$^+$)   &   &  & m1a.h09 \\
2499.22 \it10   & (1,2)$^+$& 2500.3 \it4& 2$^+$   &	6.8 \it4 & 5.7 \it3 & m1a.h11 \\										
                &        & 2507.8 \it10 &         &	1.6 \it3 &           & \\						
                &        & 2518.1 \it6  & 4$^+$   &	2.9 \it2 & 3.7 \it3 & m1a.i13 \\
2538.7 \it7     & (2$^+$)& 2536.4 \it8  & 2$^+$	  & 1.5 \it2 & 2.9 \it3 & m1a.h09 \\
                &        & 2546.9 \it10	& 6$^+$   &	0.7 \it2 & 3.3 \it3 & m2b.h09 \\
                &        & 2568.4 \it6  & 6$^+$   &	1.3 \it2 & 4.5 \it4 & m2a.h09 \\
                &        & 2578.4 \it8  &         &	3.1 \it8 &           & \\						
                &        & 2581.6 \it10 & 2$^+$	  & 3.5 \it8 & 9.4 \it4 & m1a.h09 \\
2594.73 \it20	& ($^+$) & 2594.9 \it5  & 2$^+$   &	2.9 \it3 & 6.4 \it4 & m1a.h09 \\
                &        & 2607.6 \it10	& 4$^+$   & 1.6 \it2 & 4.9 \it3 & m1a.h11 \\
                &        & 2615.9 \it6  &         &	1.6 \it2 &           & \\						
2630.9 \it5     & ($^+$) & 2632.7 \it4	& 4$^+$   & 21.7 \it9& 22.0 \it7& m1a.i13h11 \\
2644.3 \it7     &        & 2643.1 \it5  & 4$^+$   &	2.5 \it3 & 3.4 \it4 & m1a.h11i13 \\
2656.9	\it5	&        & 2657.1 \it3  & 2$^+$   & 13.0 \it5& 32.4 \it10& m1a.h09 \\
                &        & 2666.7 \it10 & 4$^+$   &	4.0 \it4 & 4.7 \it5 & m1a.h11 \\
2674.56 \it18   & (1),2$^+$ & 2673.9 \it10 &  2$^+$ & 4.3 \it6 & 4.1 \it6 &m1a.ii \\						
                &        & 2679.6 \it8  & (1$^-$) &	6.5 \it7 & 22.2 \it9 &sw.h09i \\					
                &        &          or  & (2$^+$+4$^+$)   &          &  & m1a.h09 \\
                &        & 2695.5 \it10 & 2$^+$   &  0.8 \it1 & 1.6 \it2 & sw.h09 \\		
			    &        & 2708.6 \it10 & 6$^+$   &  1.0 \it1 & 2.3 \it2 & m2a.h09 \\		
                &        & 2726.4 \it4  &  0$^+$  & 12.4 \it6& 4.3 \it7 & sw.fi\\		
                &        & 2734.0 \it4  & 2$^+$   &	10.8 \it5 & 18.4 \it10 & m1a.h09 \\
2750.43 \it19   &        & 2750.3 \it4  & 2$^+$   &	8.4 \it9 & 14.0 \it18 & m1a.h09 \\		
2758.7 \it5     & ($^+$) & 2757.2 \it4  & 0$^+$   &	15.8 \it10& 6.1 \it9& sw.h09i\\				
2761.96 \it21	&        & 2762.5 \it9  &         & 1.1 \it4 &           & \\						
2769   \it7     &        & 2771.8 \it4  & 4$^+$   &	7.5 \it4 & 25.5 \it11& m1a.h11 \\		
2782.4 \it5     & ($^+$) & 2781.6 \it6  & (6$^+$) &	4.2 \it3 & 7.1 \it7 & m2a.h09 \\		
                &        & 2799.5 \it4  & 2$^+$   &	4.0 \it3 & 9.8 \it11& m1a.h09 \\		
                &        & 2808.4 \it6  & (4$^+$) &	1.6 \it3 & 4.7 \it10& m2a.h11 \\
2822.7 \it5	    & 1$^-$	 & 2822.6 \it6  &         &	2.6 \it4 & 2.1 \it8& \\
2829.6 \it7	    & ($^+$) & 2828.5 \it5  &  2$^+$  &	3.7 \it4 & 4.6 \it5 & m1a.ii\\
                &        & 2857.0 \it5  & 4$^+$   &	3.4 \it3 & 5.5 \it6 & m1a.ii \\
                &        & 2870.4 \it10 & 4$^+$	  & 1.8 \it3 & 2.2 \it3 & m1a.ii \\
2878.8 \it4	    & 2$^+$,3& 2877.2 \it10	&         & 2.0 \it3 &		   & \\
2886		    &        & 2888.2 \it4  &  0$^+$  & 9.3 \it5 & 4.5 \it4 & sw.h11i\\
2909.6 \it5	    &	     & 2909.4 \it8  &  2$^+$  &	3.0 \it5 & 6.7 \it12& m1a.h09\\
2913.4 \it7	    &        & 2914.5 \it5  &  0$^+$  &10.9 \it6 & 4.8 \it9& sw.h09i\\
2934.6 \it11    &        & 2933.1 \it8  & 2$^+$	  & 2.3 \it3 & 6.3 \it7 & m1a.h09 \\
                &        & 2953.2 \it10	& 4$^+$   &	1.2 \it3 & 3.8 \it9& m1a.h11 \\
2961.7          &        & 2959.6 \it8  & 4$^+$	  & 4.3 \it4 & 4.3 \it8& m2a.h11 \\
2964.3 \it5	    & 2$^+$	 & 2965.8 \it20 &         &	0.5 \it3 &           & \\
2985.9 \it5	    & 1($^+$) & 2985.8 \it7  & 2$^+$  & 1.1 \it2 & 1.8 \it3 & m1a.h09 \\
                &        & 2998.3.2 \it9  & (1$^-$,2$^+$) & 0.8 \it2 & 0.5 \it2 & \\
3011.9 \it5	    & 2$^+$,3$^+$ &3012.9 \it6&  2$^+$ &6.6 \it4 & 9.9 \it6 & m1a.h09 \\
3029.2 \it6	    &        & 3029.5 \it4  &   2$^+$  &5.6 \it4 & 10.0 \it6 & m1a.h09\\
                &        & 3041.7 \it8  & (2$^+$) &	1.7 \it3 & 3.4 \it2 & m1a.h09 \\
                &        & 3053.3 \it10 & (6$^+$) &	1.1 \it2 & 1.8 \it3 & m2a.h09 \\
3060.0 \it4     & 2$^+$,3& 3061.5 \it8  & 4$^+$	  & 1.3 \it2 & 0.9 \it2 & m1a.h11i \\
3080.0 \it6     &        & 3079.2 \it6  & (6$^+$) &	2.3 \it3 & 2.4 \it2 & m2a.h09 \\
                &        & 3100.0 \it4  & 2$^+$	  & 5.1 \it4 & 10.7 \it6 & m1a.h09 \\
                &        & 3105.6 \it6  & 4$^+$	  & 2.9 \it4 & 2.7 \it3 & m1a.h11 \\
3118.5 \it15    &        & 3117.4 \it3  & 2$^+$	  & 6.0 \it3 & 11.5 \it5 & m1a.h09 \\
                &        & 3127.1 \it4  & 3$^-$   &	2.4 \it3 & 2.5 \it2 & m2a.h09\\
3141.5 \it7     &        & 3143.5 \it6  & 2$^+$	  & 0.9 \it11 & 1.7 \it3 & m1a.h09 \\
                &		 & 3145.2 \it16 &         &	1.7 \it11 &           & \\						
3150.8 \it7     & ($^+$) & 3150.4 \it20 & 4$^+$	  & 0.1 \it2 & 1.7 \it2 & m1a.h11 \\
3160.8 \it7	    &  1$^-$ & 3158.4 \it10 & 1$^-$	  & 0.4 \it2 & 1.6 \it2 & sw.ii \\
                &        & 3162.2 \it5  & 4$^+$	  & 1.9 \it3 & 1.9 \it3 & m1a.i13 \\
3171.1 \it7     &        & 3172.3 \it4  & (1$^-$,2$^+$) &	2.3 \it2 & 2.6 \it3 & \\
                &        & 3181.3 \it4  & 2$^+$	  & 4.6 \it3 & 8.3 \it4 & m1a.h09 \\
3195.4 \it6     &        & 3195.9 \it3  & 2$^+$	  & 4.6 \it6 & 6.6 \it6 & m1a.h09\\
3200.8 \it6	    & 2$^+$,3& 3200.2 \it18 & (2$^+$) & 2.0 \it6  & 0.6 \it4 & m2a.ih09\\
                &        & 3215.9 \it15 & (2$^+$) &	1.2 \it4 & 1.3 \it2 & m1a.h09\\
                &        & 3223.3 \it3  & 0$^+$	  &10.9 \it5 & 3.3 \it3 & sw.h09i\\
3234.5 \it5     &        & 3233.7 \it4  & 0$^+$   &	5.2 \it3 & 1.6 \it2 & sw.h09i\\
                &        & 3242.1 \it6	& 3$^-$	  & 1.7 \it3 & 1.4 \it2 & sw.h09 \\
                &        & 3256.6 \it4  & 2$^+$	  & 6.6 \it3 & 12.8 \it6 & m1a.h09 \\
3263.8 \it7     &        & 3265.6 \it8  & 2$^+$   &	3.9 \it2 & 4.3 \it3 & m1a.h09\\
                &        & 3276.5 \it7  & 2$^+$   &	2.7 \it3 & 2.7 \it3 & m2a.h09i\\
                &        & 3282.9 \it8  & 0$^+$   &19.5 \it5 & 6.3 \it4 & sw.fi \\
3287.9 \it5	    &        & 3288.1 \it7  & (4$^+$) & 4.3 \it10 & 3.1 \it5 & sw.ii \\
                &        & 3302.0 \it6  & 2$^+$   &	2.2 \it2 & 1.6 \it3 & m1a.ii \\
                &        & 3309.9 \it5  & 2$^+$   &	2.7 \it4 & 2.1 \it3 & m1a.ii \\
                &        & 3315.7 \it14	& 2$^+$   &	1.0 \it4 & 1.2 \it3 & sw.h09i \\
			    &        & 3327.5 \it10	& (6$^+$) & 0.3 \it2 & 0.5 \it2 & sw.h09\\
                &        & 3334.1 \it5  & 2$^+$   &	2.7 \it3 & 2.5 \it3 & sw.h09i\\
                &        & 3344.5 \it4  & (0$^+$  &	8.4 \it4 & 4.8 \it5 & sw.ih09\\
                &        &              & +4$^+$)&          & 2.0 \it5 & m1a.h11 \\
                &        & 3365.9 \it15 & 0$^+$   &	0.5 \it2 & 0.3 \it2 & sw.h09i\\
                &        & 3373.4 \it9  & 2$^+$   &	4.0 \it3 & 5.2 \it4 & m1a.h09 \\
			    &        & 3380.4 \it15 & (6$^+$) & 0.1 \it1 & 0.3 \it3 & m2b.ii\\
                &        & 3388.6 \it10 & (0$^+$) &	1.1 \it2 & 0.3 \it2 & sw.h09i \\
			    &        & 3395.5 \it4  & (4$^+$) &	0.5 \it3 & 1.0 \it3 & sw.ii\\
                &        & 3400.2 \it9  & 0$^+$   &	2.7 \it3 & 1.3 \it3 & sw.h11i\\
3411.7 \it5	    &	     & 3412.1 \it11	& 2$^+$   &	0.9 \it2 & 0.9 \it2 & sw.h09i \\
                &        & 3422.1 \it10 & 4$^+$	  & 1.1 \it2 & 1.3 \it2 & sw.hi \\
                &        & 3431.8 \it8  & 0$^+$   &	11.2 \it4& 4.2 \it3 & sw.h09i\\
3436.4 \it5     & ($^+$) & 3438.8 \it9  & (2$^+$) & 2.3 \it3 & 2.0 \it3 & m1a.ii \\
3448.8 \it5     & ($^+$) & 3447.8 \it9  & 2$^+$   &	2.1 \it2 & 1.9 \it2 & sw.ii \\
                &        & 3457.0 \it12 & 4$^+$   &	1.1 \it2 & 1.1 \it3 & m1a.ii\\						
                &        & 3463.8 \it9  & 2$^+$	  & 3.2 \it3 & 2.2 \it3 & m2a.ih \\
3469.8 \it7     &        & 3472.2 \it18 & (4$^+$) & 0.9 \it3 & 1.1 \it3 & sw.h09i\\
                &        & 3478.2 \it10 & 3$^-$	  & 3.1 \it3 & 3.4 \it3 & sw.ih \\
                &        & 3484.7 \it22	& (4$^+$) &	1.0 \it3 & 1.1 \it3 & sw.ii \\
                &        & 3490.4 \it11 & (4$^+$) &	2.7 \it4 & 3.6 \it3 & sw.ii \\
                &        & 3496.8 \it11 &  (2$^+$) & 1.4 \it2 & 2.2 \it2 &m1a.h09+0.8 \\
                &        & 3508.8 \it9  & 1$^-$   &	 0.8 \it2 & 2.6 \it5 & sw.h11 \\
			    &        & 3512.5 \it9  & 4$^+$   &	1.2 \it2 & 1.5 \it3 &m1a.ii \\
                &        & 3524.4 \it7  & 2$^+$	  & 3.0 \it2 & 3.0 \it3 & m1a.ii \\
3534.8 \it6	    & ($^+$) & 3534.1 \it6  & 2$^+$   &	4.0 \it2 & 3.1 \it3 & m1a.ii	\\
                &        & 3546.2 \it7  &  0$^+$  &	2.2 \it2 & 1.4 \it3 & sw.h11i\\
                &        & 3558.5 \it12 & 4$^+$   &	0.8 \it2 & 1.7 \it3 & sw.ii \\
3570.7 \it6	    &        & 3569.6 \it7  & 0$^+$   & 3.0 \it3 & 1.2 \it2 & sw.h09i\\
3576.8 \it7	    & 1	     & 3577.1 \it9  & 4$^+$   &	4.3 \it5 & 10.9 \it10& m1a.h11\\
                &        & 3582.9 \it7  & 3$^-$   &	6.2 \it5 & 6.6 \it7& sw.h09 \\
                &        & 3590.8 \it10 & (6$^+$) &	1.6 \it2 & 2.1 \it5 & m2a.ih09\\
                &        & 3603.1 \it10 & 2$^+$   &	1.2 \it2 & 3.8 \it3 & m1a.h09 \\
                &        & 3616.6 \it8  & 0$^+$   & 10.8 \it4 & 4.2 \it3 & sw.h09i\\
3626.9 \it6	    & ($^+$) & 3626.4 \it8  & 0$^+$   &	24.6 \it5 & 8.6 \it4 & sw.fi\\
                &        & 3635.6 \it4  & 2$^+$   &	2.3 \it3 & 2.2 \it3 & m1a.ii \\
                &        & 3641.7 \it8  & 0$^+$   &	4.4 \it4 & 1.3 \it3 & sw.h09i\\
                &        & 3651.1 \it9  & 1$^-$   &	1.9 \it2 & 3.2 \it3 & sw.ih\\
3655.4 \it8	    & 1,2$^+$& 3657.9 \it4  & 2$^+$   &	1.2 \it2 & 2.1 \it3 & m1a.hi \\
3663.3 \it10    &        & 3665.7 \it10 & (6$^+$) &	1.2 \it2 & 1.6 \it3 & m1a.ii \\
                &        & 3676.3 \it9	& 2$^+$   &	2.6 \it3 & 2.5 \it4 & m1a.ii\\
                &        & 3681.3 \it13 & 2$^+$   &	1.1 \it3 & 3.8 \it4 & m1a.hi \\
                &        & 3691.7 \it8  & 0$^+$   &	22.2 \it6& 7.4 \it5 & sw.fi\\
                &        & 3699.7 \it14 & 4$^+$   &	1.3 \it3 & 1.8 \it4 & m1a.h11\\
                &        & 3706.5 \it10 & 2$^+$	  & 1.9 \it3 & 2.5 \it3 & m1a.h09 \\
                &        & 3712.0 \it4  & 1$^-$   &	0.4 \it2 & 1.2 \it3 & sw.ii \\
                &        & 3721.2 \it11 & 2$^+$   &	1.1 \it2 & 1.5 \it3 & m1a.h09 \\
                &        & 3737.9 \it11 & 0$^+$   &	2.9 \it7 & 0.9 \it2 & sw.h09i\\
                &        & 3741.8 \it15 & 4$^+$	  & 1.7 \it7 & 1.6 \it3 & sw.h09i \\
                &        & 3761.8 \it6  & 4$^+$   &	2.5 \it2 & 2.3 \it2 & m1a.ii\\
                &        & 3777.0 \it6  & 2$^+$   &	2.9 \it2 & 1.9 \it2 & m1a.ii \\
                &        & 3784.5 \it4	& 4$^+$	  & 0.2 \it2 & 0.8 \it2 & sw.h09 \\
3794.6 \it10    &        & 3790.0 \it9  & 2$^+$   &	1.2 \it2 & 1.4 \it2 & m1a.h09 \\
                &        & 3802.5 \it10 & (2$^+$) &	1.0 \it2 & 0.3 \it2 & m1a.h09\\
                &        & 3811.1 \it10 & 2$^+$   &	1.2 \it2 & 1.0 \it2 & m1a.ih09 \\
3819.8 \it7	    & 1$^-$  & 3819.2 \it7  & (0$^+$  &	2.4 \it2 & 1.5 \it2 & sw.ii \\
                &        &              & +4$^+$) &          & 0.6 \it2 & m1a.h11 \\
                &        & 3829.1 \it6  &  0$^+$  & 5.5 \it3 & 2.3 \it3 & sw.h09i\\					
3846.6 \it5	    & ($^+$) & 3848.1 \it8  &  0$^+$  &	2.8 \it3 & 2.0 \it2 & sw.h11i\\
                &        & 3853.7 \it10 &  4$^+$  &	0.9 \it3 & 0.5 \it2 & sw.h09i\\
                &        & 3865.2 \it13 & 6$^+$   &	0.6 \it1 & 0.9 \it2 & sw.ff \\
                &        & 3876.1 \it6  & 0$^+$   &	5.6 \it4 & 2.1 \it3 & sw.h09i\\
                &        & 3881.9 \it11 & 4$^+$   &	1.8 \it3 & 2.2 \it3 & sw.hi \\
                &        & 3892.5 \it10 & 4$^+$	  & 1.0 \it2 & 1.1 \it2 & sw.h09i \\
                &        & 3908.7 \it28 & 4$^+$   &	0.2 \it1 & 1.1 \it2 & sw.h09 \\
3923.9 \it6     &        & 3925.9 \it11 & 2$^+$   &	0.9 \it2 & 0.9 \it2 & m1a.h09 \\
                &        & 3937.9 \it17 & 1$^-$   &	0.6 \it2 & 0.8 \it2 & sw.hi \\
3948.0 \it6	    &	     & 3946.1 \it8  & 2$^+$   &	2.2 \it2 & 3.1 \it2 & m1a.h09 \\
                &        & 3959.3 \it16	& (4$^+$) &	1.0 \it3 & 2.2 \it3 & sw.ii \\
3965.1 \it7	    &        & 3965.9 \it18 & 4$^+$   &	1.1 \it3 & 1.5 \it4 & m1a.ii \\
                &        & 3974.2 \it9	& 2$^+$   &	2.2 \it2 & 2.3 \it2 & sw.h09 \\
                &        & 3984.9 \it6  & 0$^+$   &	7.8 \it3 & 3.7 \it3 & sw.h11i\\
                &        & 3995.1 \it16 & 2$^+$   &	1.1 \it2 & 1.6 \it2 & m1a.h09 \\
                &        & 4000.5 \it4  & 2$^+$   &	1.0 \it3 & 0.7 \it2 & m2a.ih \\
4015.8 \it8     &        & 4014.1 \it9  & 3$^-$   &	1.3 \it2 & 1.2 \it2 & sw.ii \\
                &        & 4024.5 \it9  & 3$^-$   &	1.4 \it2 & 1.5 \it2 & m2a.h09\\
                &        & 4038.1 \it9  & 2$^+$   &	1.3 \it2 & 1.1 \it2 & sw.hh \\
                &        & 4049.9 \it13 & 2$^+$   &	0.8 \it2 & 0.6 \it2 & sw.hi \\
                &        & 4058.8 \it4  & (2$^+$) &	0.5 \it2 & 1.0 \it3 & m1a.h09\\
                &        & 4066.1 \it6	& 2$^+$   &	3.8 \it2 & 2.3 \it4 & m1a.ii\\
                &        & 4086.4 \it7  & 4$^+$   & 2.2 \it2 & 1.6 \it2 & m1a.ii\\
                &        & 4097.6 \it8  & 2$^+$   &	1.7 \it2 & 1.2 \it2 & sw.hi \\
                &        & 4114.3 \it9  & 2$^+,4^+$	 & 1.8 \it2 & 1.4 \it2 & m1a.hi \\
                &        & 4123.5 \it11 & (2$^+$) &	1.2 \it2 & 0.9 \it2 & m1a.h09 \\
                &        & 4153.0 \it13 & (4$^+$) &	1.2 \it4 & 1.1 \it3 & m1a.ii \\
                &        & 4159.8 \it21 & 2$^+$   &	1.0 \it3 & 1.6 \it4 & m1a.h09 \\
                &        & 4167.6 \it14	& 2$^+$   &	1.4 \it3 & 1.6 \it3 & m1a.h09 \\
                &        & 4176.8 \it11 & 4$^+$   &	1.2 \it2 & 1.3 \it2 & sw.ii \\
                &        & 4191.3 \it9  & 4$^+$   &	1.5 \it2 & 1.6 \it2 & sw.ii \\
                &        & 4206.3 \it8  & 3$^-$   &	1.7 \it2 & 1.3 \it2 & sw.ii \\
                &        & 4220.5 \it8  & 0$^+$   & 2.7 \it3 & 1.8 \it3 & sw.h09i\\
                &        & 4228.0 \it11 & 2$^+$   &	1.7 \it3 & 1.2 \it3 & m1a.ii \\
                &        & 4250.1 \it10 & 4$^+$   &	1.2 \it2 & 1.4 \it3 & sw.ii\\
                &        & 4258.1 \it6  & 0$^+$   &	3.6 \it3 & 2.5 \it3 & sw.h09i\\
                &        & 4272.4 \it9  & 3$^-$   &	1.9 \it2 & 1.7 \it3 & sw.ii \\
                &        & 4281.7 \it34 & 4$^+$   &	0.4 \it2 & 1.8 \it3 & sw.h09 \\
\hline \\
\end{longtable*}

\newcolumntype{d}{D{.}{.}{3}}
\begin{table*}[]
\caption{\label{Tab:bands} \normalsize{The sequences of  states in $^{158}$Gd
 which can be treated as rotational bands as follows from  the CHUCK fit, the (p,t) cross
 sections and the moment of inertia (values of  $J/\hbar^2$ are given). }}
\begin{ruledtabular}
\begin{tabular}{cccccccccc}\\
$K^{\pi}$ & $0^+$ & $1^+$ & $2^+$ & $3^+$ & $4^+$ & $5^+$ & $6^+$ & $7^+$
& MoI [MeV$^{-1}$] \\
\hline\\
0$^+$ & 0.0     &&    79.5  &&  261.5   &&   539   &&   37.5 \\
2$^+$ &        && 1187.2   & 1265.5 & 1358.5 & 1481.4 & 1623.5 && 37.3\\
0$^+$ &1196.2  && 1259.9   &&  1406.7  && 1635.5  && 46.9\\
0$^+$ &1452.4  && 1517.5   &&  1667.4  &&         && 45.9   \\
0$^+$ &1743.2  && 1791.8   &&  1901.6  &&         && 61.1 \\
2$^+$ &        && 2026.3   && 2202.5   && 2471.4  && 41.9 \\
0$^+$ &1957.3  && 2035.6   &&&&&&  38.3 \\
0$^+$ &1977.6  && 2056.5   &&&&&&  38.0   \\
2$^+$ &        && 2084.0  && 2230.3 &&&&  47.9\\
2$^+$ &        && 2089.3 & 2153.5 & 2239.3 && 2471.3 && 46.7 \\
2$^+$ &        && 2098.0 &        & 2249.0 && 2481.8 && 46.3 \\
2$^+$ &        && 2218.7 && 2383.5 &&&& 42.5 \\
2$^+$ &        && 2260.3 && 2413.3 &&&& 45.7 \\
2$^+$ &        && 2283.4 && 2463.3 &&&& 38.9 \\
0$^+$ & 2276.6 && 2344.2 && 2493.8 && 2708.6 && 44.4 \\
2$^+$ &        && 2354.8 && 2518.1 && 2781.6 && 42.9 \\
0$^+$ & 2437.2 && 2500.3 && 2643.1 &&&& 47.5 \\
2$^+$ &        && 2657.1 && 2771.8 &&&& 61.1 \\
2$^+$ &        && 2734.0 && 2857.0 && 3053.3 && 56.9 \\
2$^+$ &        && 2750.3 && 2870.4 && 3053.3 && 58.3 \\
0$^+$ & 2726.4 && 2799.5 && 2959.6 &&&& 41.1 \\
0$^+$ & 2757.2 && 2825.3 && 2959.6 &&&& 42.1 \\
2$^+$ &        && 2909.4 && 3061.5 && 3327.5 && 42.2 \\
2$^+$ &        && 2933.1 && 3105.4 && 3380.4 && 40.6 \\
0$^+$ & 2914.5 && 2985.8 && 3150.4 &&&& 42.1 \\
2$^+$ &        && 3029.5 && 3162.4 && 3380.4 && 52.7 \\
2$^+$ &        && 3100.0 && 3288.4 && 3590.5&& 37.2 \\
2$^+$ &        && 3181.3 && 3344.5 && 3590.5 && 42.9 \\
2$^+$ &        && 3256.6 && 3422.1 && 3665.8 && 42.5 \\
2$^+$ &        && 3265.6 && 3395.5 &&&& 53.9 \\
2$^+$ &        && 3276.5 && 3457.0 &&&& 38.8 \\
0$^+$ & 3223.3 && 3302.0 && 3484.7 &&&& 38.1 \\
0$^+$ & 3233.7 && 3309.5 && 3490.4 &&&& 39.2 \\
0$^+$ & 3282.9 && 3334.1 && 3457.0 &&&& 58.6 \\
2$^+$ &        && 3373.4 && 3512.5 &&&& 50.3 \\
0$^+$ & 3344.5 && 3412.1 &&&&&& 44.4 \\
0$^+$ & 3400.2 && 3447.8 && 3558.5 &&&& 62.9 \\
0$^+$ & 3431.8 && 3524.4 && 3741.9 &&&& 32.4 \\
2$^+$ &        && 3534.1 && 3699.7 &&&& 42.3 \\
2$^+$ &        && 3603.1 && 3761.8 &&&& 44.1 \\
0$^+$ & 3569.6 && 3635.6 && 3784.6 &&&& 45.4 \\
0$^+$ & 3616.6 && 3676.3 && 3819.2 &&&& 50.3 \\
2$^+$ &        && 3681.3 && 3853.7 &&&& 40.6 \\
0$^+$ & 3626.4 && 3706.6 && 3892.5 &&&& 37.4 \\
0$^+$ & 3641.7 && 3721.2 &&&&        && 37.7 \\
0$^+$ & 3691.7 && 3777.0 && 3965.9 &&&& 35.2 \\
2$^+$ &        && 3790.0 && 3959.3 &&&& 41.4 \\
0$^+$ & 3737.9 && 3811.2 &&       &&&&  40.9 \\
0$^+$ & 3829.1 && 3925.9 && 4153.0  &&&& 31.1 \\
0$^+$ & 3848.2 && 3925.9 &&         &&&& 38.61 \\
0$^+$ & 3876.1 && 3946.1 && 4114.3 &&&& 42.5 \\
2$^+$ &        && 3974.2 && 4176.8 &&&& 39.2 \\
2$^+$ &        && 3995.1 && 4191.3 &&&& 38.5 \\
0$^+$ & 3984.9 && 4066.0 && 4250.1 &&&& 37.0 \\
\vspace{0mm}\\
1$^-$ && 0977.2 & 1023.7 & 1041.6 & 1159.0 & 1176.5 & 1371.9 & 1390.6 & 51.8 \\
1$^-$ && 1263.5 && 1402.9 && 1638.3 &&& 35.9 \\
1$^-$ && 1856.3 & 1894.6 & 1978.1 &&&&& 41.1 \\
1$^-$ && 3158.4 && 3242.1 && 3395.5 &&& 59.7 \\
1$^-$ && 3508.8 && 3582.9 &&&&& 67.5 \\
1$^-$ && 3937.4 && 4014.1 &&&&& 65.2 \\
\end{tabular}
\end{ruledtabular}
\end{table*}
\section{Discussion of results}

\subsection{Collective rotation bands and moments of inertia in $^{158}$Gd}

Since the success of the Bohr-Mottelson generalized rotation model \cite{Bohr98},
many advanced approaches to the nuclear rotation  have been developed.
They are reviewed, for example, in the book of D.J.Row \cite{Row10}.
For the purposes of this subsection, we use the simplest model \cite{Bohr98},
which  successfully describes rotational  bands in strongly deformed nuclei,
such as $^{158}$Gd.

After the assignment of spins to all excited states,
sequences of states which show the
characteristics of a rotational band structure can be distinguished.
An identification of the states attributed to rotational bands was
made on the basis of the following conditions:

{\bf \emph{i}}) the angular distribution for a state as  a band member
candidate is fitted by the DWBA calculations for the spin value
that is necessary to put this state  into the band;

{\bf \emph{ii}}) the transfer cross section in the (p,t) reaction to the
states in the potential band has to decrease with increasing spin;

\begin{figure}
\includegraphics[width=0.48\textwidth,clip]{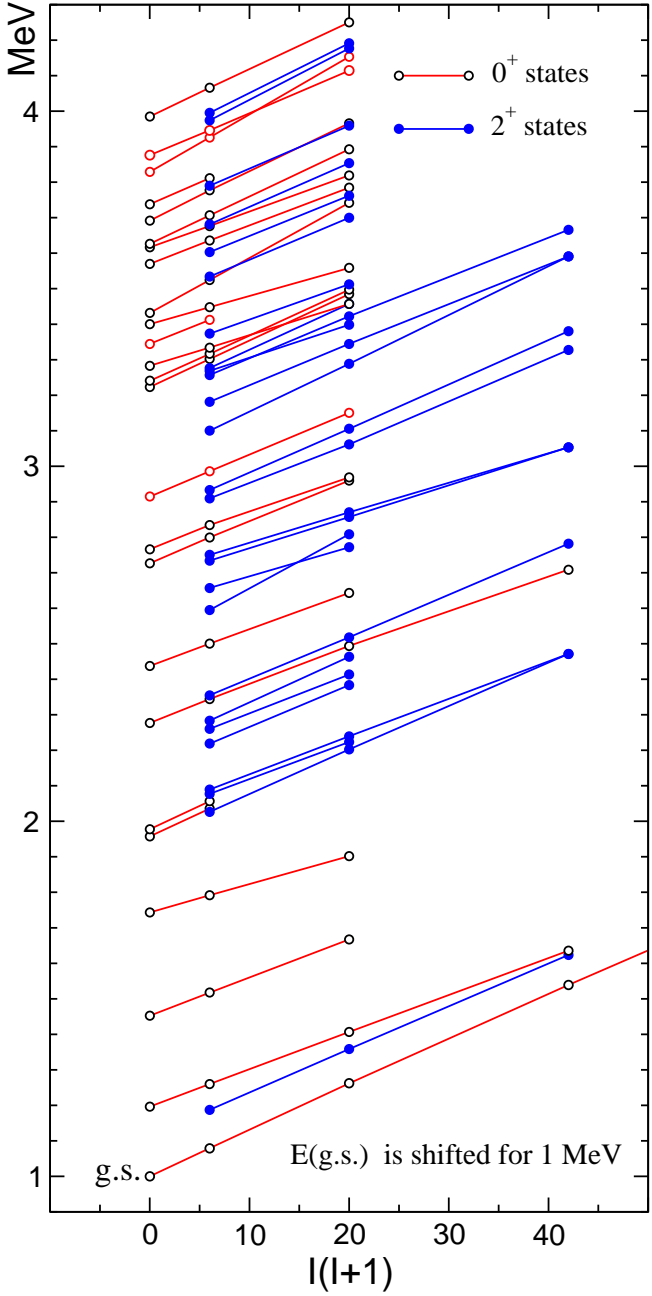}
\hspace{-5mm}
\caption{\label{bands}
    Collective bands based on the 0$^+$
   and 2$^+$ excited states in $^{158}$Gd
   as assigned from the DWBA fit of the angular distributions
   found from the (p,t) reaction as functions of the spin variable
   $I(I+1)$.}

\end{figure}

{\bf \emph{iii}}) the energies of the states in the band can be approximately fitted
 by the expression for a rotational band
\begin{eqnarray}
E_{\rm rot} = E_K +
\frac{\hbar^2}{2J}[I(I+1)-K(K+1)] ~.
\label{enrot}
\end{eqnarray}
Thereby, a rotational band is  unambiguously
identified by the energy $E_K$ of
a band head
with a $K$
quantum number  -  the projection of
 the total angular momentum onto
 the symmetry axis for a given band head,
 and $J$, which is the moment of inertia (MoI)
  (below  in text we use MoI for $J$ in
   Eq.~(\ref{enrot})).
   Collective bands identified
in such a way are shown in Fig.~\ref{bands} and
the energies $E_{\rm rot}$  are
listed in Table \ref{Tab:bands}.
  This procedure can be
  justified by the fact
that some
sequences meeting the above criteria are already known
from gamma ray spectroscopy to be rotational bands,
so other similar sequences are very probably rotational
bands too.
Nevertheless,  additional information
(on E2 transitions at least) is needed
to definitely confirm these assignments.

Within a rotational band, its members share
 almost the
same MoI, i.e.
  only small, relatively smooth
variations of
the MoI value with
increasing spin  may {occur, and this is emphasized by the straight lines
  in Fig.~\ref{bands}.
 The moments of inertia  calculated through the slopes  of
 these lines are listed in Table~\ref{Tab:bands}.

It can be expected that the
 MoI reflects the intrinsic structure of the
rotational band, for which the pairing interaction is important.
Fig.~\ref{MoI}   demonstrates
that the  MoI magnitudes
for  most excited states in  $^{158}$Gd
are larger  than that of the  g.s..
They are located in a region limited by the  g.s. value
and that of the first excited bands known from previous studies.
Most of them  have values close to that of the ground state MoI
equal  approximately to 37.5 MeV$^{-1}$.
According to Ref.~\cite{Bohr98}, vibrational bands have MoI that
are typically a few percent larger than that of the  g.s. band.
More than half of the bands based on 0$^+$ states reveal just this property.
The bands with a significantly larger MoI are supposedly based
on two-phonon states or having even more complicated phonon structure.
The two-quasiparticle states with spins 2$^+$
and higher can also be detected in the spectra, although the cross
section for their excitation is expected to be weak.
 Due to the blocking effect, rotational bands built on such
states may exhibit MoI 30 - 50\% larger than that for
the ground state band \cite{Bohr98}.
Some bands have MoI lower than those of the g.s.
Two of them are 2$^+$ state bands, their MoI are only about 1\%
lower than that of the ground state.  One band head at 1187.2 keV
is a $\gamma$ - vibrational state.
Five of the heads of such bands are 0$^+$ states with the MoI between 31.1 and 37.4 MeV$^{-1}$,
they are located above 3400 keV, much higher than
twice the energy gap.  Their structure is intriguing.

  \begin{figure}[h]
  \hspace{4.5mm}
\includegraphics[width=0.44\textwidth,clip]{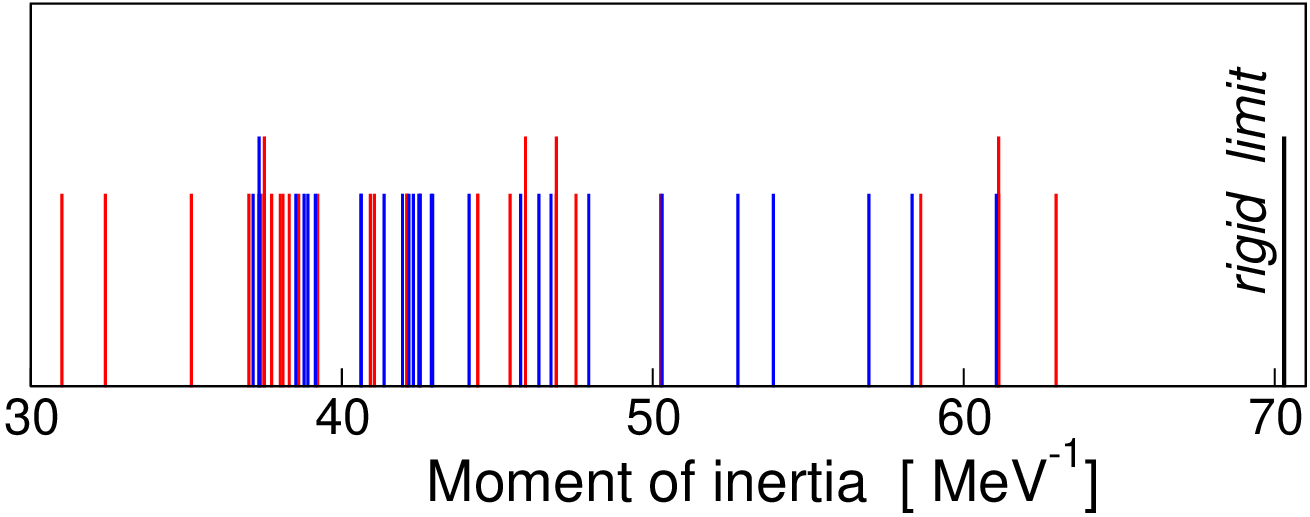}

\includegraphics[width=0.48\textwidth,clip]{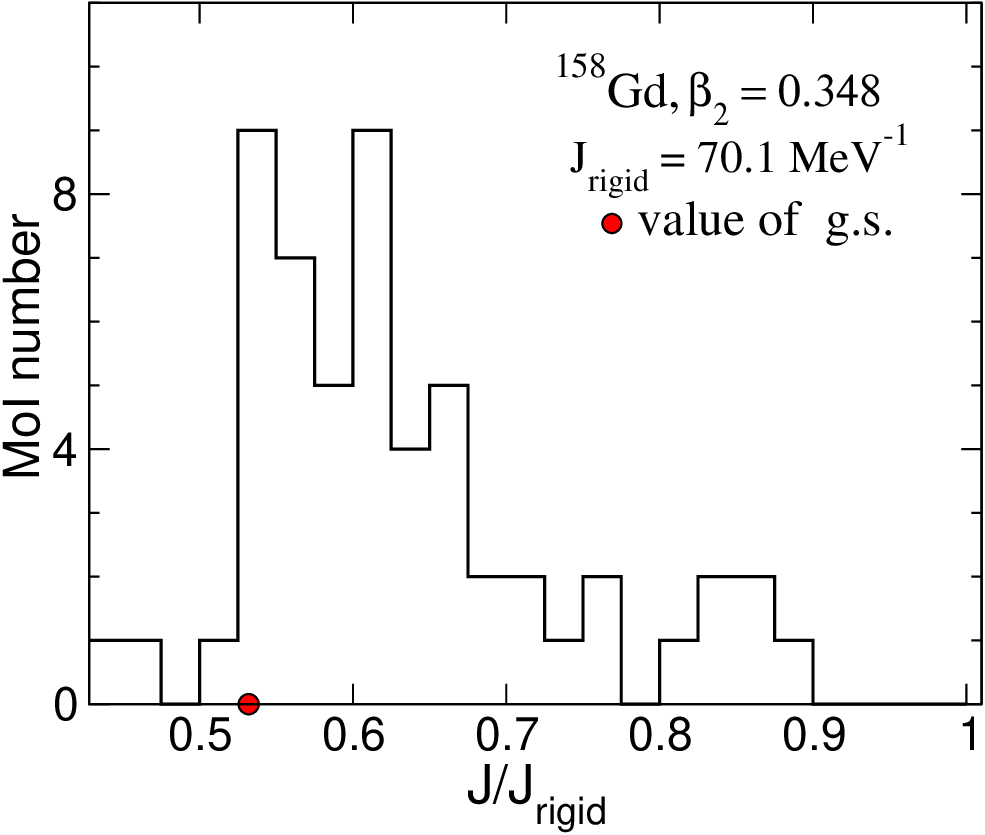}
\caption{\label{MoI}
{\bf \textsl{Top}:} The moments of inertia (MoI)
as determined from  the assigned collective bands.
The moments of inertia
for 0$^+$  and 2$^+$ states are shown  by
red and blue lines, respectively.
 Larger height represents the data known from previous studies.  \\
{\bf \textsl{Bottom}:}
Distribution of number of the MoI values  versus the
dimensionless values $J/J_{rigid}$. The value of $J_{rigid}$ is evaluated
according to Eq.~(\ref{srMoI}). A sampling interval is 0.025. }
\end{figure}

It is well known that the nucleus in the lowest excited states has
 MoI values  which
do not exceed approximately 50\%
of the moment of inertia of a rigid rotator with the same  nuclear mass.
 A part of the
nucleons  of the nucleus
is not involved in the rotational motion due to the effect of
the nucleon pairing, which
leads to superfluid properties of nuclei
in the ground and lower excited states.
The moment of inertia for a statistically equilibrium rotation \cite{LLv5}
 can be approximated  as the rigid body limit
 \cite{Sitenko14},
\begin{eqnarray}
  \frac{J_{rigid}}{\hbar^2} = \frac{2}{5}\frac{mA^{5/3}r_0^2}{\hbar^2}(1+0.32\beta_2)~.
  \label{srMoI}
\end{eqnarray}
 where a shape of spheroid  with the deformation $\beta_2$
was assumed  for the nucleus.
For $^{158}$Gd,  the rigid-rotator MoI value  (\ref{srMoI})
is about 70 MeV$^{-1}$.
 The standard deformation parameter $\beta^{}_2$  describing mainly the nuclear
 shape is another  important characteristic affecting the MoI magnitude.
  Due to the pairing effect, one can expect that the
  MoI magnitude deviates much from the rigid-rotator limit
   (\ref{srMoI}), namely, the MoI decreases by
  about 44\%.
  Thus,
  the two factors - nuclear deformation and pairing -
  and, in addition,
the centrifugal stretching
   can be considered as the main reasons of
  a significant increase of  the MoI
  with increasing excitation energy, as compared
  to the ground state value.
 The largest value of the  MoI is equal to
  63 MeV$^{-1}$, that is, almost 90\% of the rigid-body limit (\ref{srMoI}).
  The distribution of the MoI values relative to the rigid-rotator value
  (\ref{srMoI}) is shown in Fig.~\ref{MoI}.

\subsection{Statistical analysis of the 0$^+$ and 2$^+$ state sequences
     and possible $K$ symmetry breaking.}


Sequences of states observed in  the extended excitation energy
interval in $^{158}$Gd are considered to be long enough to perform statistical
analysis even for one nucleus, see Table \ref{tab_number}.  The
  present analysis is triggered by  the publication of Paar and Vorkapi
\cite{Paa90}, which  is devoted to the
investigation of  effects of  the exact $K$
quantum number
on the fluctuation properties of the energy spectra for 0$^+$ and 2$^+$
states  in the SU(3) limit of the IBM.
The $\Delta_3$ statistics \cite{Dys63} was used
to obtain information about the long-range correlations of level spacings.
In Ref.~\cite{Paa90}, the $\Delta_3$  statistics  for  the
pure sequence of the $0^+$
levels is close to the Wigner (chaotic) behavior while for  the
mixed sequence of all $2^+$
levels is close to the Poisson (regular) behavior
(see  also Ref.~\cite{Gom11}).
The $\Delta_3$  statistics
with the fixed $K$ sequences  $(I =2,K =0)$ and $(I =2,K =2)$
return back to the Wigner distribution.

The sequences of states  considered above as
rotational bands  look basically
long enough  to carry out  the statistical
analysis  both for $K$ mixed sequences of 2$^+$  and $4^+$ states
and, separately, for the sub-sequences with  $(I =2,K =0)$ and $(I =2,K =2)$
 as well as for  those with
$(I =4,K =0)$, $(I =4,K =2)$ and $(I =4,K =4)$.
The number of levels in all  such sequences is shown
in Table \ref{tab_number}.

\newcolumntype{d}{D{.}{.}{3}}
\begin{table} [h]
\begin{ruledtabular}
\begin{tabular}{ccccc}
~$I/K$~  &  ~All~ &  ~$K = 0$~ & $K = 2$~ & ~$K = 4$~\\
\hline
\vspace{0.1mm}\\
~0$^+$~  &  37  &&&\\
2$^+$    &  100 & 37 & 63 &\\
4$^+$    &  90  & 37 & 28 & 25 \\
\vspace{0.1mm}\\
All      &  227 & 74 & 92 & 25 \\
\end{tabular}
\end{ruledtabular}
\caption{\normalsize {Number of levels included in the statistical analysis.
They can be compared with the corresponding numbers at study of
the isospin symmetry breaking in $^{30}$P: 102 of all levels, 69 for $T=0$ and 33 for $T=1$}.}
\label{tab_number}
\end{table}

The  nearest neighbor-spacing distributions (NNSD) \cite{Wig65,Lev18}
are applied
to investigate the fluctuation properties of
short-range correlations of the experimental spectra.
The  NNSDs  are fitted  by using the linear
Wigner-Dyson  approximation   LWD
with
one parameter $w$ \cite{Mag18},
\begin{equation}
p^{}_{\rm LWD}(s)=[a(w)+b(w) s]~\exp\left[-a(w) s - \frac{b(w)}{2} s^2\right],
\label{pslinWD1}
\end{equation}
where
\begin{equation}
a=\sqrt{\pi}~w~e^{w^2_{}}~\mbox{erfc}(w)~,\qquad
b=\frac{\pi}{2}~e^{2w^2_{}}~\mbox{erfc}^2_{}(w)~.
\label{abzpar}
\end{equation}
 $\mbox{erfc}(w)=1-\mbox{erf}(w)$, $\mbox{erf}(w)$ is the error function.\\

The  LWD
allows to obtain
information on the quantitative measure of the Poisson
regular and Wigner chaotic  contributions, separately,
in contrast to the heuristic Brody 
 parameterization \cite{Bro73} with a
fitting parameter which  has not, in this respect,
a clearly defined meaning.
Results of fitting for two angular momenta, 0$^+$ and 2$^+$
are shown in Fig.~\ref{fig10} and in Table \ref{table4}.
 For calculations of the experimental NNSDs,
  simple polynomials of low powers
  were used
  for fitting well the staircase cumulative level density obtained
  from experiments
  to get the so called unfolding (uniformed dimensionless) energy levels, see
Ref.~\cite{Lev18} for details.
 NNSDs  for the spin 2$^+$ with  the fixed
angular-momentum  projections $K=0$ (c) and $2$ (d) have a Poisson-like
structure, similar to the NNSD for 0$^+$ state (a).
The  NNSD for  the spin 2$^+$ without fixing  (``all K'')
the angular momentum
 projection is shifted to the Wigner distribution.
 Fig.~\ref{fig11} and Table \ref{table4} show  the results of the
 analysis for the spin $I=4$.
The NNSDs   have the  Poisson-like structure for all sequences, except
for the ($I=4$, $K=2$) one which demonstrates a noticeable shift towards
the Wigner  distribution.


\newcolumntype{d}{D{.}{.}{3}}
\begin{table}[h]
\begin{ruledtabular}
\begin{tabular}{cccccc}
~$I^\pi$~  &~$K$~ & ~ $a$~ &~ $b$ ~ &  $w$ ~&~ $\chi^2_2$~\\
\hline
\vspace{0.1mm}\\
~0$^+$     & $0$  &~ 98.1~ & ~1.9~ &~ 5.04~& ~11.6\%~\\
~ 2$^+$ & all $K$ & 58.1   & 41.9   & 0.66   &14.9\% \\
~ 2$^+$ & 0       & 98.2   & 1.8   & 5.2   &9.8\%\\
~2$^+$ & 2$^+$    & 90.8   & 9.2   & 2.2   &13.2\% \\
\vspace{0.1mm}\\
~4$^+$& all $K$   &~ 82.3~ & ~17.7~ &~ 1.4~ &~17.7\%\\
~ 4$^+$ &~0~      & 98.2   & 1.8   & 5.1   &8.9\% \\
~ 4$^+$ & ~2~     & 76.2   & 23.8   & 1.1   &16.5\%\\
~4$^+$ &~ 4~      & 98.2   & 1.8   & 5.2   &14.0\% \\
\end{tabular}
\end{ruledtabular}

\caption{\normalsize {
    Parameters $a$ and $b$ are  one-parameter
     LWD approximation  (\ref{pslinWD1})
  within the Wigner-Dyson theory
for the  excited states 0$^+$,  for 2$^+$ with all $K$,
fixed $K=0$ and $K=2$ and for  4$^+$ with all $K$,
fixed $K=0$, $K=2$ and $K=4$ in $^{158}$Gd.
NNSD parameters
for Poisson and Wigner contributions $a$ and $b$ are
reduced to total 100\%.  Large
$w$ ($w=\infty$) corresponds
to Poisson and small
$w$ ($w=0$) is related to  the Wigner limits.
 The standard accuracies found by $\chi^2_i$
 of  the least-squares fittings are given in percent.
 A sampling interval is 0.2.}}
\label{table4}
\end{table}

\begin{figure*}
\includegraphics[width=0.68\textwidth,clip]{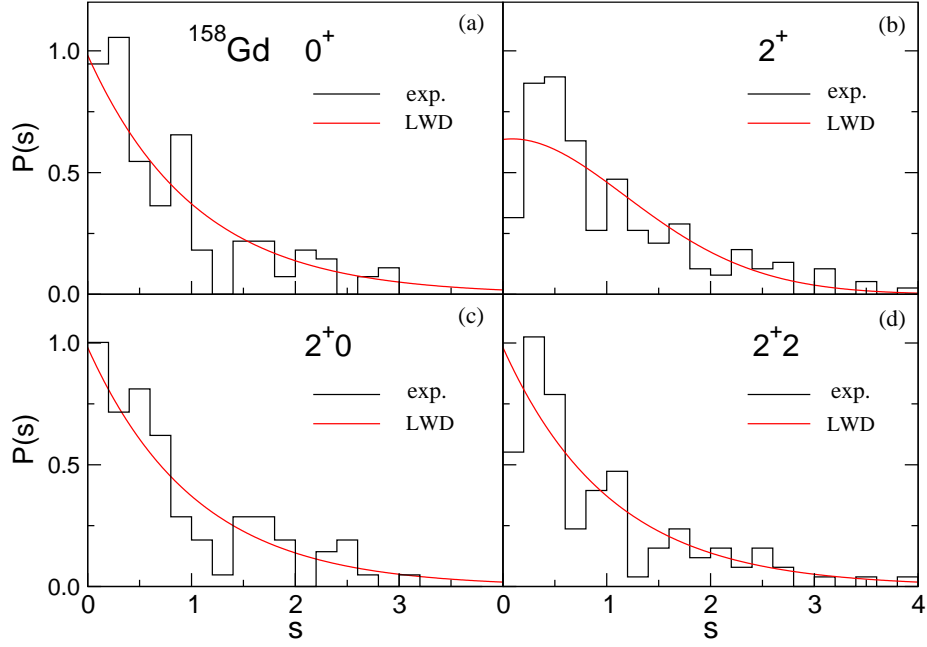}
\vspace{-0.1cm}
\caption{{
 The  nearest neighbor-spacing distributions  for 0$^+$ states (a) and
    for 2$^+$ states with all projections $K$ (b), fixed $K=0$ (c) and $K=2$ (d)
projections. }}
\label{fig10}
\end{figure*}

\begin{figure*}
\includegraphics[width=0.68\textwidth,clip]{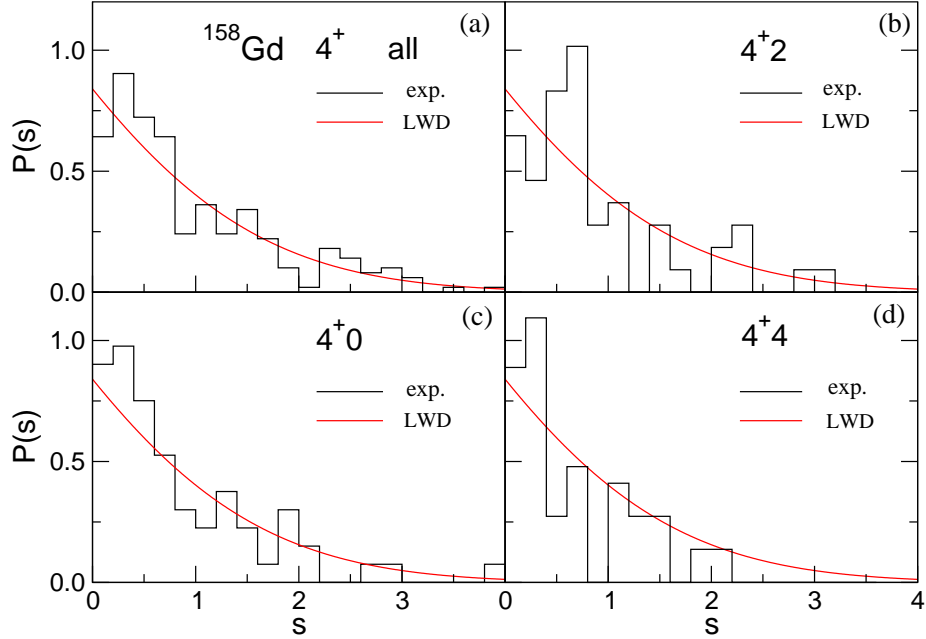}
\vspace{-0.1cm}
\caption{{
   The nearest neighbor-spacing distributions  for  4$^+$ states
     with all projections $K$ (a), fixed $K=0$ (b), $K=2$ (c) and $K=4$ (d)
projections.}}
\label{fig11}
\end{figure*}

Joining  the sets of 0$^+$ states in the rare  earth and
actinide nuclei, which became available from the rich data
obtained during the last decades
\cite{Lev94,Wir04,Lev09,Lev13,Lev15,Spi13,Mey06,Buc06,Bet09,Ili10,Spi18,Lev19},
 demonstrate intermediate statistics between
 the Wigner and Poisson limits \cite{Lev18}.
 As shown in Ref.~\cite{Lev18} the level spacing distributions for
 the collective 0$^+$,  2$^+$ and 4$^+$  states
 mixing all $K$ in the actinide nuclei were found to be  gradually shifted
 to the Poisson limit with increasing   the spin \cite{Lev18}.
 In $^{158}$Gd we observe
  a different behavior: practically  a pure Poisson statistics
 for 0$^+$ states and an essential shift to  the
 Wigner distribution for 2$^+$ states with all $K$.
 In the case of 4$^+$ states we  find the level spacing distribution  close
 to the Poisson limit for a sequence that includes all $K$ and for
 the sub-sequences ($I=4, K=0$) and ($I=4, K=4$).   However,
   the level spacing distribution
 for the sub-sequence $(I=4, K=2)$ demonstrates again a slight shift towards
  the Wigner limit.

The experimental results for the fixed $K$ projections in the case of
 2$^+$ as well
 as of 4$^+$ states differ from the calculations  performed
 in Ref.~\cite{Paa90}.
 These results cannot be  compared directly since the
 $\Delta_3$  statistics analysis has been performed for long correlations
 in Ref.~\cite{Paa90}.
 However, from a general point of view,
  having a good quantum number $K$ one should expect a shift to the Wigner
 distribution in the sub-space of the fixed $K$ value for a given angular
 momentum $I$ with respect to the case of accounting for all $K$
 \cite{BM98,BM12,LM19}. This is  because of decreasing of the single-valued
 integral-motion numbers (conservation laws) due to a breaking of the
 axial symmetry in the sub-space versus
 its presence in the complete space. In such a sub-space, one finds
 more a system disordering or chaos
\footnote{{This interpretation of the $K$-breaking effect differs from another more
 discussed  in literature \cite{zelev96}.
 Alternatively, we may think of the
 $K$ symmetry breaking as an effect of violating the axial symmetry when $K$
 is not a good quantum numbers due to an additional interaction, e.g., the
 $\gamma$ deformation above the alongation $\beta^{}_2$ considered here.}}.
The arguments for this interpretation  of the
$K$-breaking are working well for the NNSDs in the
case of actinide nuclei \cite{Mag18,LM19}.
The present results for $^{158}$Gd differ from those in
Ref.~\cite{Lev18,Mag18} and are not so clearly understood.
Only the case of ($I$=4,$K=2$) can be considered
as supporting to some extent  this interpretation,
 see  Fig.~\ref{fig11} (b) and Fig.~\ref {fig11} (a).
Its NNSD is between the  Wigner and Poisson limits, i.e.
 is not so pronounced as in actinide nuclei \cite{Mag18,LM19}.
And Wigner's contribution to the NNSD in this case  (b) is much less
  than Poisson one.
As for the remaining  sub-sequences with  $(I =2,K =0)$ and $(I =2,K =2)$
 as well as for sub-sequences with  $(I =4,K =0)$ and $(I =4,K =4)$,
 they are strongly shifted to a regular Poisson distribution.
 Although the number of levels used in the analysis is limited,
 this  affects only the accuracy of determining the  Wigner and Poisson
 contributions.
Such behavior might be interpreted as a $K$ symmetry breaking
 when $K$ is a good quantum number.
That is a subject  for further study in forthcoming work.

The effects of
another symmetry breaking on the level statistics
 were
studied  experimentally in $^{26}$Al \cite{Mit88} and $^{30}$P \cite{Shr00,Shr04}.
 Statistical analyses have been performed taking into account
 the isospin quantum number $T$.
 The experimental distributions P(s) occurred to be
 equally far from  Wigner  and Poisson  limits
  as for  the ($I^\pi;T$)  sequences, and for the ($I^\pi$) sequences
   when  the isospin quantum number $T$ is ignored.
  The reason for this behavior may be that,
  although sequences of different $T$ are not correlated if isospin
  describes an exact symmetry,
even a small breaking of isospin symmetry
may lead to similar fluctuation properties for ($I^\pi;T$)
and the ($I^\pi$) sequences \cite{Gom11}.
According to theoretical calculations \cite{Dys62} and \cite{Pan81}
in that case spectral fluctuations may be nearly independent of $T$.
Probably similar situation is met also  in the $^{158}$Gd nucleus
 when $K$ is a good quantum number.

A puzzle is remaining  why the NNSD for the mixed 2$^+$ sequence demonstrates a shift
to the Wigner distribution,  cf. (b) with (a) in Fig.~\ref{fig10}.
The present analysis includes the states excited in the (p,t) reaction.
According to previous studies \cite{Lev09,Lev13,Lev15,Buc06,Spi18},
the multiple 0$^+$ states excited in the (p,t) reaction are  found to be
collective.
This
 is perhaps not the case for  2$^+$ states; the
excitations of states of another nature are not excluded,
though with  a smaller cross section.
To verify this assumption and to see
how these states
 can influence on the results of statistical analysis,
  non-collective 2$^+$ states from the compilation ~\cite{Nic17},
not observed in the present (p,t) experiment, were  included in
the analysed sequence.
The obtained P(s) turned out to be additionally shifted to the Wigner
distribution in comparison with  that shown in Fig.~\ref{fig10} (b).
The presence of non-collective states in the  sequence of 2$^+$ states
can be probably one of the reasons of such observed NNSD for 2$^+$ states
shifted to the Wigner limit as compared to that for 0$^+$ and 4$^+$ states.
Non-collective levels are  probably
absent in the $(I =2,K =0)$ sequence and present
in the $(I =2,K =2)$ one what is reflected in the Wigner-Poisson contributions.\\

\section{IBM calculations}
The structure of $^{158}$Gd was investigated in the framework of the Interacting Boson Model.
 The traditional version of the IBM \cite{Iach87} does not make any distinction
 between protons and neutrons and uses only $\it{s}$ and $\it{d}$ bosons
 (with angular momentum L=0 and 2, respectively) as the main ingredients
 to describe the low-lying positive-parity states  of even-even nuclei.
 Several other versions have been proposed over the years that include
 the addition of several other type of bosons, like $\it{p}$, $\it{f}$
 and $\it{g}$ (with angular momentum  L= 1, 3 and 4, respectively).
 In the last 20 years, new and detailed data have been measured with
 the (p,t) reaction and a considerable amount of states, especially 0$^+$,
 have been found. One of the interpretation of this increased number
 of 0$^+$ excitations was given by the IBM using the $\it{spdf}$ version
 of the model. The reason is that by coupling of two  negative-parity
 bosons the model produces additional {\it K}$^\pi$=0$^+$ states which
 have a N$_{pf}$=2 configuration. Such calculations have been performed
 in Refs. \cite{Zamfir01,Lev09,Lev13,Lev15}  and have shown a rather
 good reproduction of the overall trend of electromagnetic and hadronic
 observables. This interpretation involves an increased contribution of
 the octupole degree of freedom in the low-lying structure of nuclei,
 which is  in  disagreement with  a prediction of other theoretical models,
 for example, the Quasiparticle Phonon Model  (QPM).
 The QPM indicates
 a moderate contribution of the octupole components in their wave
 functions while
  gives an increased weight of the pairing correlations \cite{Lo05}.
 Therefore, one needs experimental data  concerning
 different type of observables in order to
 test  properly the two predictions.
 The case of $^{158}$Gd is one of the most promising examples for the following
 reason.  In the rare-earth region,
 this is the only nucleus
 that  has
 information both
  from the (p,t) transfer reaction and from a dedicated neutron inelastic
  scattering experiment aimed at measuring the lifetimes of the new 0$^+$
  excitations in (p,t) \cite{Lesh07}. Together with known transition
  probabilities of the lowest octupole states, we have a very fertile
  testing ground of the IBM predictions.

Therefore, we have performed calculations in the  \textit{spdf} IBM-1 framework
using the Extended Consistent Q-formalism (ECQF) \cite{Casten01}.
Although the equations employed by the model have been given in several papers,
e.g.  Refs.~\cite{Zam01,Zam03,Zamfir01,Scholten01},
 we briefly list them again below. The usual Hamiltonian is given by:
\begin{eqnarray}
\hat{H}_{spdf}=\mathrm{\epsilon}_{d} \hat{n}_{d}+\mathrm{\epsilon}_{p}
\hat{n}_{p}+\mathrm{\epsilon}_{f} \hat{n}_{f} + \mathrm{\kappa}(\hat{Q}_{spdf}\cdot
\hat{Q}_{spdf})^{(0)}\nonumber\\
+\alpha \hat{D}^{\dagger}_{spdf}\cdot \hat{D}_{spdf}~,\label{eq1}
\end{eqnarray}
where \(\epsilon_{d}\), \(\epsilon_{p}\), and \(\epsilon_{f}\)
are the boson energies and \(\hat{n}_{p}\), \(\hat{n}_{d}\),
and \(\hat{n}_{f}\) are the boson number operators.
We mention that one of the ingredients that was shown
to improve the transfer calculations, namely the inclusion
of the octupole term in the Hamiltonian \cite{Lev13, Lev15},
was omitted in the present calculations since we preferred
to maintain the form of the Hamiltonian given in Ref. \cite{Zamfir01}.
 $ \hat{D}_{spdf}$ is introduced in the Hamiltonian in order
 to connect states with no (\(pf\)) content with those
 having \((pf)^2\) components, and it has a very small strength
 as shown in Table \ref{Table1}. The form of this operator
 is taken as  earlier, see Refs. \cite{Zam01,Zam03},
\begin{eqnarray}
\hat{D}_{spdf}=-2\sqrt{2}[{p}^{\dagger}\tilde{d}+{d}^{\dagger}\tilde{p}]^{(1)}
+\sqrt{5}[{s}^{\dagger}\tilde{p}+{p}^{\dagger}\tilde{s}]^{(1)}\\\nonumber
+\sqrt{7}[{d}^{\dagger}\tilde{f}+{f}^{\dagger}\tilde{d}]^{(1)}~.\label{eq2}
\end{eqnarray}
For the quadrupole operator one has \cite{Kusnezov01}:
\begin{eqnarray}
\hat{Q}_{spdf}=\hat{Q}_{sd}+\hat{Q}_{pf}=\nonumber\\
(\hat{s}^{\dagger}\tilde{d}+\hat{d}^{\dagger}\hat{s})^{(2)}+\chi^{(2)}_{sd}(\hat{d}^{\dagger}\tilde{d})^{(2)}
+\frac{3\sqrt{7}}{5}[(p^{\dagger}\tilde{f}+f^{\dagger}\tilde{p})]^{(2)}\nonumber\\
- \frac{9\sqrt{3}}{10}(p^{\dagger}\tilde{p})^{(2)}-\frac{3\sqrt{42}}{10}(f^{\dagger}\tilde{f})^{(2)}\label{eq3}
\end{eqnarray}

The quadrupole electromagnetic transition operator is  defined by
\begin{eqnarray}
\hat{T}(E2)=e_{2} \hat{Q}_{spdf}\label{eq4}
\end{eqnarray}
where \(e_{2}\) represents the boson effective charge.

\begin{table}[b]
  \caption{\label{Table1}  The
    $spdf$-IBM parameters used in the present calculations.
The parameters of the Hamiltonian are taken from  Ref.~\cite{Zamfir01},
while the others are determined from a fitting procedure on the
corresponding experimental data.}
\begin{ruledtabular}
\begin{tabular}{cccc}
\multicolumn{2}{c} {Parameters} & \multicolumn{2}{c} {Nucleus}\\
\cline{3-4}\\
                       &                          &$^{158}$Gd    & $^{160}$Gd \\
\hline
                       & $\epsilon_d$ (MeV)   & 0.315           & 0.213   \\
                       & $\epsilon_p$ (MeV)   & 4.0               & 4.0   \\
  Hamiltonian  & $\epsilon_f $ (MeV)   & 0.95            & 1.3            \\
                       & $\kappa$ (MeV)        & -0.02            & -0.02  \\
                       & $\chi_{sd}$                & -0.91       & -0.53  \\
                       & $\alpha$ (MeV)         & 0.0005	      & 0.0005 \\
\hline
                       & $e_2$ (eb)                 & 0.132	       & 0.132  \\
EM transition  & $e_1$  (eb$^{1/2}$)   & 0.053	        & 0.053             \\
 operators       & $\chi_{sp}$                & 1.07             & 1.07      \\
                       & $\chi_{df}$                 &  -0.55    & -0.55     \\
\hline
               & $\alpha_\nu$ (mb/sr)          &\multicolumn{2}{c} {0.008}   \\
Transfer operator  & $\alpha_p$ (mb/sr)      &  \multicolumn{2}{c} {4.22}    \\
                & $\alpha_f$ (mb/sr)              &  \multicolumn{2}{c} {-0.4} \\
                \end{tabular}
\end{ruledtabular}
\end{table}

\begin{figure*}
\includegraphics[width=0.8\textwidth,clip]{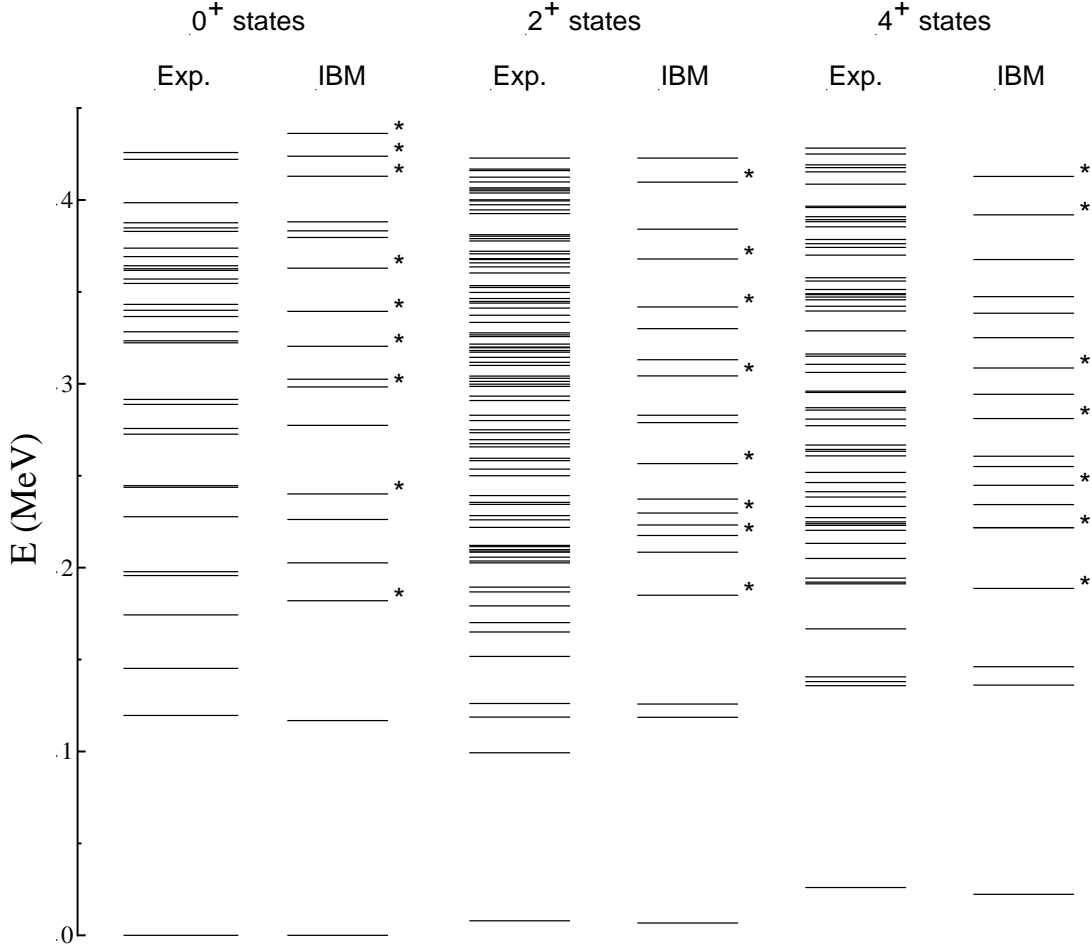}
\caption{\label{Fig_024} Comparison between the experimental
and $spdf$-IBM calculations for the 0$^+$, 2$^+$, and 4$^+$ states
up to 4.3 MeV. The levels with a double-octupole character in the IBM are marked with a star.}
\end{figure*}

\begin{figure*}
\centering
\includegraphics[width=15cm, angle=0]{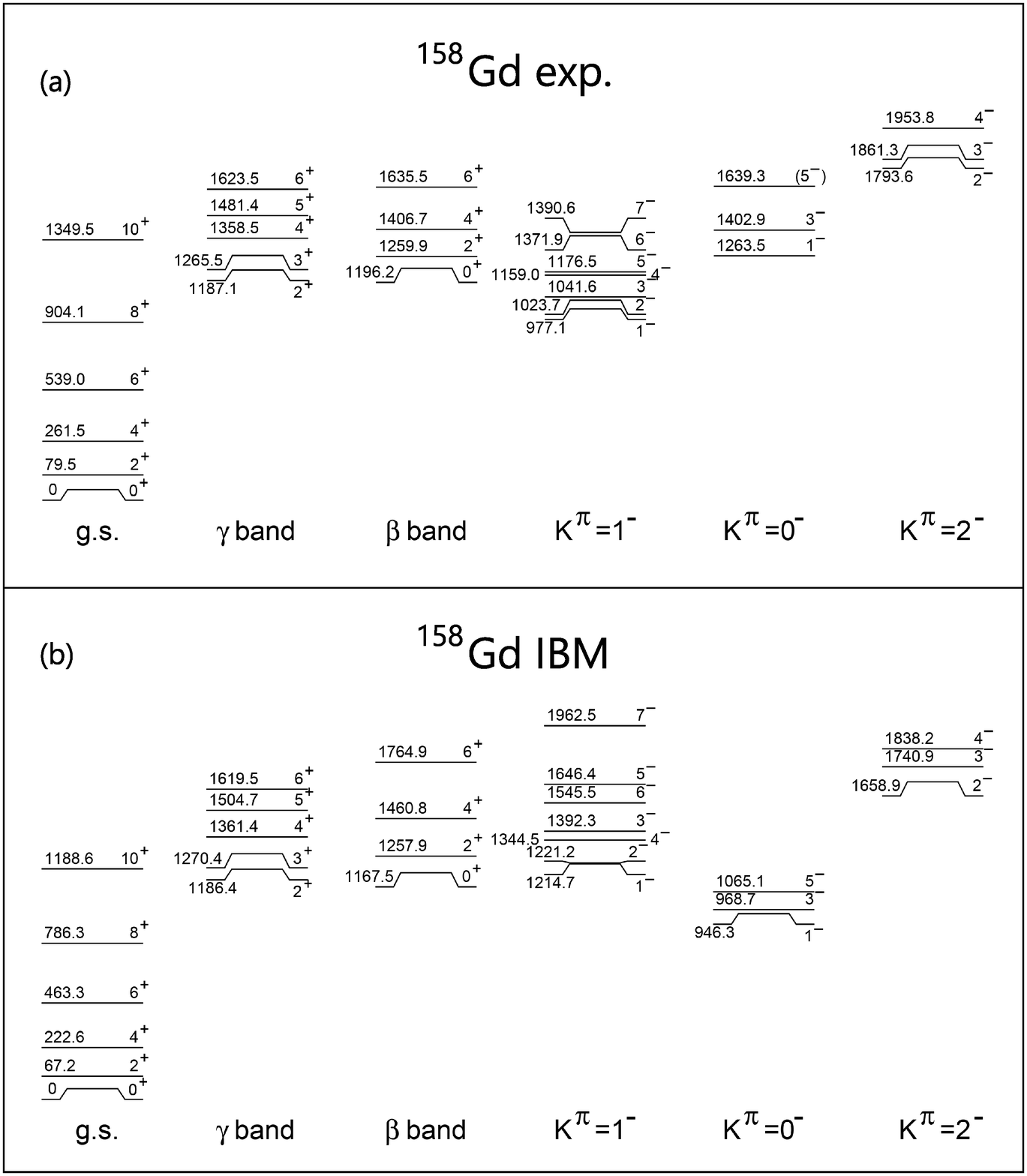}
\caption{\label{fig1} Experimental   (a) and
  $spdf$-IBM   (b)
level scheme of $^{158}$Gd. The g.s., $\gamma$, and $\beta$ band are shown
for the positive-parity states, while the K$^\pi$=0$^-$, 1$^-$ and 2$^-$
octupole bands are presented for the negative-parity levels.}
\end{figure*}

Since the IBM
 yields the increased octupole correlations
in the structure of even-even nuclei,
it is essential to calculate the \(E1\) transition strengths and to compare
the results with the experimental values.  For the \(E1\) operator
in the IBM
 one has

\begin{eqnarray}
\hat{T}(E1)=e_{1}[\chi_{sp}^{(1)}({s}^{\dagger}\tilde{p}+{p}^{\dagger}\tilde{s})^{(1)}+
({p}^{\dagger}\tilde{d}+{d}^{\dagger}\tilde{p})^{(1)}\nonumber\\
+\chi_{df}^{(1)}({d}^{\dagger}\tilde{f}+{f}^{\dagger}\tilde{d})^{(1)}]~,\label{eq5}
\end{eqnarray}
where \(e_{1}\) is the effective charge for the \(E1\) transitions and \(\chi_{sp}^{(1)}\)
 and \(\chi_{df}^{(1)}\) are two model parameters.

 The final equation  which we need
 is the one for the transfer operator. Previously,
only the last term in Eq. (11) was used \cite{Scholten01}, but recent successful
calculations \cite{Lev13,Lev15} have shown that it is imperative to include also
at least one term related to the negative-parity  bosons

\begin{eqnarray}
\hat{P}^{(0)}_{\nu}=(\alpha_{p}\hat{n}_{p}+\alpha_{f}\hat{n}_{f})\hat{s}+\nonumber\\
+\alpha_{\nu} \left (\Omega_{\nu}-N_{\nu}-\frac{N_{\nu}}{N} \hat{n}_{d}\right)^{\frac{1}{2}} \left(\frac{N_{\nu}+1}{N+1}\right)^{\frac{1}{2}} \hat{s}\label{eq6}
\end{eqnarray}
where \(\Omega_{\nu}\) is the pair degeneracy of neutron
shells, \(N_{\nu}\)
is the number of neutron pairs, \(N\) is the total number of bosons,
and \(\alpha_{p}\), \(\alpha_{f}\),
and \(\alpha_{\nu}\) are constant parameters.

Schematic $\it{spdf}$-IBM calculations  have been performed
in Ref. \cite{Zamfir01}
shortly after limited data on 0$^+$ states in $^{158}$Gd were obtained in the
(p,t)  experiment \cite{Lesh02}.
With more data on hand, we proceed to investigate
not only the distribution in energy of the 0$^+$ states, but also
the  detailed structure of $^{158}$Gd, including the energies of
the low-lying levels,
the transition probabilities in the first bands and the distribution in
transfer intensity of the  0$^+$ states up to 4.5 MeV.
To perform the calculations we
employed the OCTUPOLE  code \cite{Kusnezov02} to diagonalize the Hamiltonian
in Eq.(\ref{eq1}). Up to three negative-parity bosons were allowed in
the calculations and the parameters of the Hamiltonian were taken from
 Ref.~\cite{Zamfir01},
while the ones for the transition and transfer operators were fitted to the available
experimental information. The IBM parameters are summarized in
Table \ref{Table1}.

\begin{figure}
\centering
\includegraphics[width=8.5cm, angle=0]{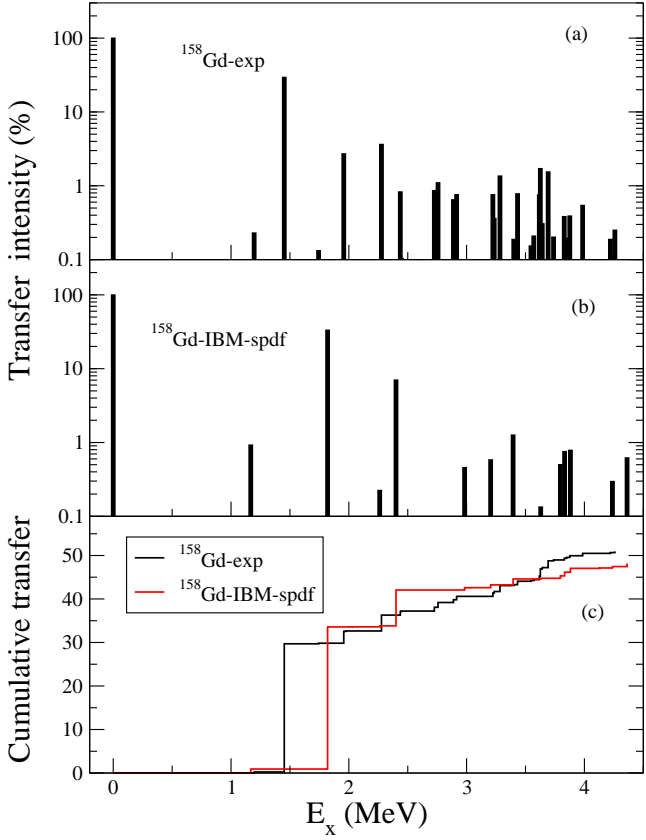}
\caption{\label{fig2} Comparison between the experimental
   (a)
  and $spdf$-IBM
  calculations   (b) for the transfer intensity
  in $^{158}$Gd. Cumulative
  strength as a function of energy is given in   (c)
  for experiment (black) and calculations (red).}
\end{figure}

The authors of Ref. \cite{Zamfir01} have presented a comparison of the experimental
 energy levels with the corresponding ones calculated in the $spdf$-IBM framework.
 Their work concentrated mainly on the reproduction of the 0$^+$ states and it
 was for the first time when the model predicted an increased number of 0$^+$
 levels, close to the experimentally observed one. The contribution of
 the octupole degree of freedom was crucial, the model
 describing  twelve 0$^+$ states
 up to around 3.5 MeV, where the experimental data were available at that time.
In Fig. \ref{Fig_024} we  present the complete results of the IBM calculations
for the 0$^+$, 2$^+$, and 4$^+$ states up to 4.3 MeV in comparison with
the values obtained in the present experiment. It is clear that the experiment
has revealed a greater number of states that can be produced by the IBM,
irrespective of spin. The experiment provides 36   (0$^+$), 95 (2$^+$),
and 64 (4$^+$) states, while the IBM gives only 17 (0$^+$),
20 (2$^+$), and 19 (4$^+$)  states.
Some of these levels
  having a double-octupole character
are marked with
a star in Fig. \ref{Fig_024}.
It is clear that this version of the IBM is satisfactorily describing
the low energy part of the spectra, but is not so successful in
describing the spectra at higher excitation energies, and a
more complicate version should be used or other models
have to be considered
in order to elucidate the structure of $^{158}$Gd.

In Fig. \ref{fig1} we compare the energy levels of the lowest positive- and
negative-parity bands, using the data from the latest evaluation in
ENSDF \cite{Nic17}.
One observes a rather good reproduction of the experimental data, especially
of the positive-parity states. For the negative-parity levels, the calculations
 show a band order with K$^\pi$=0$^-$, 1$^-$ and 2$^-$, while in the experiment
the order is K$^\pi$=1$^-$, 0$^-$ and 2$^-$. This effect was previously noticed
in the IBM calculations \cite{Cottle01} and it was related to the fractional
filling of the proton and neutron valence shells. The ordering can be improved
in the IBM by introducing another term in the calculations that will lower
 the  K$^\pi$=1$^-$ band in energy \cite{Cottle01}. However, since we try
 to keep the calculations as close as possible to the ones in Ref. \cite{Zamfir01},
 this term was not included in the  Hamiltonian,
 Eq.~(\ref{eq1}).

 The results
  for the transfer intensity calculated in the IBM
  by using Eq.(\ref{eq6})
  are  compared with the experimental data in Fig. \ref{fig2}.
As noted above, the IBM does not reproduce the number of 0$^+$ states
 obtained in the present experiment:  17 excited  0$^+$ levels in the
IBM calculations are  found
 versus the 36 experimental 0$^+$ excitations
in the energy region  under the consideration.
 It is clear that some of the observed 0$^+$ excitations
 are having a two quasi-particle nature and are, therefore,
 outside  of the model space.
  Thus, detailed microscopic calculations are needed to
  reproduce the structure
 of all these states. Nevertheless, we look also at the transfer
 intensity produced
 by the IBM model in order to see how much
 the observed strength may have
 a collective origin.
 In Fig.~\ref{fig2},
  (a) and (b), we present the experimental
and calculated transfer strength, respectively. One can see that the IBM does give
a reasonable reproduction of the experimental data for the transfer intensity.
The first excited 0$^+$ state has 0.2$\%$ of the ground strength in the experiment
and 0.9$\%$ in the calculations, while the second excited 0$^+$ state has
about 30$\%$ and 34$\%$ in experiment and calculations, respectively.
 For higher-lying excitations,  one obtains
about 20$\%$ in the experiment,
and amount to about 14$\%$ in the IBM. The distribution of the transfer strength
is better illustrated in Fig. \ref{fig2} (c), where we compare the experimental
and calculated cumulative transfer. It is clear that the model reproduces
the experimental data up to about 3.5 MeV, and starts to underestimate it
at higher excitation energy. It will be interesting to obtain experimental
data for energies even higher than 4.3 MeV in future experiments to better
compare the distribution in energy and transfer strengths of the higher-lying states.

\begin{table}
\caption{\label{Table2} Experimental and calculated $E1$ and $E2$ transition
probabilities in $^{158}$Gd. The parameters of the $E1$ and $E2$ operators
are fitted to the experimental data available \cite{Nic17}.}
\begin{ruledtabular}
\begin{tabular}{cccccc}
K$^{\pi}$ & E$_{i}$ (keV) & J$_{i}$ & J$_{f}$  & Exp. (W.u.) & IBM (W.u.) \\
\hline
\multicolumn{6}{c} {$E2$ transitions}\\
\hline
g.s                  &   80    &   2$^{+}$     &       g.s.            &  198(5)      & 198     \\
\vspace{0.05cm}
                       &  261   &   4$^{+}$     &  2$^{+}_{1}$    &  290(4)      & 280     \\
\vspace{0.05cm}
                       &  904   &   8$^{+}$     &  6$^{+}_{1}$    &  330(30)    & 308     \\
\vspace{0.05cm}
                       & 1350  &   10$^{+}$   &  8$^{+}_{1}$    &  340(30)    & 304     \\
\vspace{0.05cm}
$\beta$-band  & 1196  &   0$^{+}$     &  2$^{+}_{1}$    &  1.17$^{+4.18}_{-0.13}$       & 4.79\\
\vspace{0.05cm}
                      &  1260  &   2$^{+}$     &  4$^{+}_{1}$    & 1.39(15)    & 3.25     \\
\vspace{0.05cm}
                      &            &   2$^{+}$     &  2$^{+}_{1}$    & 0.079(14)  & 0.47     \\
\vspace{0.05cm}
                      &            &   2$^{+}$     &  0$^{+}_{1}$    & 0.31(4)      & 0.97     \\
\vspace{0.05cm}
                      &  1407  &   4$^{+}$     &  2$^{+}_{\beta}$         & 456$^{+912}_{-67}$     & 180     \\
\vspace{0.05cm}
                      &            &   4$^{+}$     &  2$^{+}_{\gamma}$    & 12.8$^{+25.6}_{-1.9}$  &1.2      \\
\vspace{0.05cm}
                      &            &   4$^{+}$     &  6$^{+}_{1}$         & 3.16$^{+6.32}_{-0.46}$    & 3.66     \\
\vspace{0.05cm}
                      &            &   4$^{+}$     &  4$^{+}_{1}$        & 0.37$^{+0.74}_{-0.05}$    & 0.005     \\
\vspace{0.05cm}
                      &            &   4$^{+}$     &  2$^{+}_{1}$        & 1.32$^{+2.64}_{-0.19}$    & 1.01     \\
\vspace{0.05cm}
$\gamma$-band  & 1187  &   2$^{+}$     &  4$^{+}_{1}$    &  0.27(4)       & 0.12   \\
\vspace{0.05cm}
                            &            &  2$^{+}$     &  2$^{+}_{1}$    &  6.0(7)          & 5.65     \\
\vspace{0.05cm}
                            &            &   2$^{+}$     &  0$^{+}_{1}$   &  3.4(3)         & 2.23    \\
\vspace{0.05cm}
                           & 1266   &   3$^{+}$     &  4$^{+}_{1}$    &  1.77$^{+3.27}_{-0.19}$       & 3.29  \\
\vspace{0.05cm}
                            &            &  3$^{+}$     &  2$^{+}_{1}$    & 3.5$^{+6.47}_{-0.37}$    & 4.61     \\
\vspace{0.05cm}
                           & 1358   &   4$^{+}$     &  2$_{\gamma}$    &  113$^{+166}_{-13}$       & 99   \\
\vspace{0.05cm}
                           &            &  4$^{+}$     &  6$^{+}_{1}$          &    $ >$0.95        & 0.07     \\
\vspace{0.05cm}
                           &            &  4$^{+}$     &  4$^{+}_{1}$          & 7.3$^{+10.7}_{-0.9}$   & 6.9     \\
\vspace{0.05cm}
                           &            &  4$^{+}$     &  2$^{+}_{1}$          & 1.13$^{+1.65}_{-0.14}$  & 0.43     \\
\hline
\multicolumn{6}{c} {$E1$ transitions}\\

\hline
1$^-$    & 977     &   1$^{-}$     &  2$^{+}_{1}$    &  9.7$^{+12.7}_{-1.1}$ $\cdot$10$^{-5}$  & 5.2$\cdot$10$^{-5}$   \\
\vspace{0.05cm}
   &     &   1$^{-}$     &  0$^{+}_{1}$    &  9.8$^{+12.8}_{-1.1}$ $\cdot$10$^{-5}$       & 29.2$\cdot$10$^{-5}$   \\
\vspace{0.05cm}
 & 1042   &   3$^{-}$    &  4$^{+}_{1}$    &  2.9(8)$\cdot$10$^{-4}$                        & 3.3$\cdot$10$^{-4}$   \\
\vspace{0.05cm}
  &              &   3$^{-}$   &  2$^{+}_{1}$    & 3.3(10)$\cdot$10$^{-4}$                 & 0.8$\cdot$10$^{-4}$   \\
\vspace{0.05cm}
 & 1159   &   4$^{-}$    &  4$^{+}_{1}$    &  9.3$^{+18.6}_{-1.2}$ $\cdot$10$^{-5}$       & 12.1$\cdot$10$^{-5}$   \\
\vspace{0.05cm}
 & 1176   &   5$^{-}$    &  6$^{+}_{1}$    &  5.9$^{+6.7}_{-0.7}$ $\cdot$10$^{-4}$        & 10.9$\cdot$10$^{-4}$   \\
\vspace{0.05cm}
  &            &   5$^{-}$    &  4$^{+}_{1}$    &  7.4$^{+8.4}_{-0.8}$ $\cdot$10$^{-4}$       & 0.72$\cdot$10$^{-4}$   \\
\vspace{0.05cm}
0$^-$        & 1264  &   1$^{-}$     &  2$^{+}_{1}$    &  6.4(21)$\cdot$10$^{-3}$             & 9.5$\cdot$10$^{-3}$   \\
\vspace{0.05cm}
 &            &   1$^{-}$     &  0$^{+}_{1}$    &  3.5(12)$\cdot$10$^{-3}$                     & 4.3$\cdot$10$^{-3}$   \\
\vspace{0.05cm}
    & 1403 &   3$^{-}$    &  4$^{+}_{1}$    &  1.6(3)$\cdot$10$^{-2}$                         & 1.0$\cdot$10$^{-2}$   \\
\vspace{0.05cm}
    &          &  3$^{-}$     &  2$^{+}_{1}$    & 1.2$^{+0.2}_{-0.23}$ $\cdot$10$^{-2}$      & 0.8$\cdot$10$^{-2}$    \\
\vspace{0.05cm}
2$^-$          & 1794  &   2$^{-}$     &  3$^{+}_{\gamma}$    &  5.0$^{+10.0}_{-0.6}$ $\cdot$10$^{-5}$  & 167$\cdot$10$^{-5}$ \\
\vspace{0.05cm}
  &            &   2$^{-}$     &  2$^{+}_{\gamma}$    &  8.6$^{+17.2}_{-1.1}$ $\cdot$10$^{-5}$   & 294$\cdot$10$^{-5}$   \\
\vspace{0.05cm}
  &           &   2$^{-}$    &  2$^{+}_{1}$       &  1.8$^{+35.0}_{-0.2}$ $\cdot$10$^{-7}$       & 5410$\cdot$10$^{-7}$   \\
\end{tabular}
\end{ruledtabular}
\end{table}

 Finally, we look at the reduced matrix elements that can provide a better insight
if the relevant degrees of freedom are taken into account. For the case of $^{158}$Gd, most
of the lifetimes have been measured for the low-lying states, with both positive and
negative parity \cite{Nic17}. Therefore, an impressive amount of B(E1) and B(E2) values
is available to be compared with the theoretical
 calculations. In Table \ref{Table2}
we present the IBM results
for the $E1$ and $E2$ transition
 probabilities for the g.s, $\beta$, and $\gamma$ band, as well as for
 the  K$^\pi$=0$^-$, 1$^-$ and 2$^-$ octupole bands. The model reproduces
 the gross features of the low-lying states in $^{158}$Gd, but a closer
 inspection reveals that there are some severe discrepancies with respect
 to the experimental data.   The
 $E1$ transitions in
 the K$^\pi$=2$^-$ band are  found
 to be much stronger than in
 the experiment, although the experimental uncertainty is quite large.
 The same situation happens for the $E1$ transition from the 0$_2^+$ state
 to the 1$_1^-$. However, most of the transitions are
  obtained within
 less than a factor of five as compared to the experimental data.
 For the higher-lying 0$^+$ states, the (n,n'$\gamma$) experiment
 has revealed a low $E1$ transition strength up to around 3 MeV \cite{Lesh07}.
 Since the double-octupole states play a major role in the IBM,
 it is not surprising that many of these states are predicted with
 a relatively high $E1$ transition strength. Therefore, we conclude
 that although the low-lying structure of $^{158}$Gd is reasonable
 well reproduced by the IBM calculations, the theory does not reproduce
 in details the nature of the  higher-lying 0$^+$ states.

\subsection{Conclusion}
A  proper
study of excited
states with energies up to 4.3 MeV in the deformed nucleus $^{158}$Gd
was performed   by
a high-resolution (p,t) transfer reaction using
the Q3D spectrograph. In total, 206 excited states of positive parity
and 20 of negative parity were identified and many of
 them were observed for the first time.
The high resolution, background-free experiment allowed, in fact,
a quasi-complete determination of levels  up to excitation
energies with  a  high level density.
The collective nature of these states is provided by the selectivity
of the (p,t) reaction to the structure of the  densely populated
final states.
 To assign spin and parity to the states, angular distributions were
measured and compared to the predictions of coupled-channel
DWBA calculations. Many rotational bands built upon the
low-lying band heads excited in our experiment were
identified.  Moments of inertia calculated using energies of such
 bands are  analysed. The large  sets
 of states with the same spin-parity
allowed to carry out their statistical analysis. Such an
analysis  is performed  for the 0$^+$ and 2$^+$  states sequences
including all K-values and for well-determined projections K
of the angular momentum.  We intended to obtain confirmation
of theoretical predictions about the chaotic nature of sequences
with a well-determined projection $K$ of the angular momentum.
However, all but  one analysed NNSDs
indicate   clearly on the regular nature.
Although the number of levels used in the analysis is limited,
which affects the accuracy of  determination of  the Wigner and Poisson
 contributions, we interpreted this behavior as an indication of
 the $K$ symmetry  breaking with $K$ being a good quantum number.
More detailed analysis of such data for the rare-earth and actinide nuclei
is a subject  for further study in forthcoming work.
The structure of $^{158}$Gd was investigated in the framework
of the Interacting Boson Model using the $spdf$
 version of the model. The calculated  energies
of the low-lying levels, their transition probabilities in the
lowest bands and  their distributions in  the transfer intensity of
0$^+$ states are in rather good agreement  with the experiment.
 We found clear signatures to go beyond the simplest \textit{spdf}
 version of the IBM in describing the complete  data.
 The description of such rich experimental data by more sophisticated,
 (semi)microscopical theoretical models, is of considerable interest.

\section{Acknowledgments}

We are grateful to the members of the YMKB collaboration
for access to the $^{158}$Gd data, and thank Dr.~Deseree A. Meyer
(Brittingham) for useful discussions. We thank also
the operators at MLL for excellent beam conditions.
The work was supported in part by the Romanian project PN 18090102F2.
This work was supported  in part  also by the  budget program
"Support for the development
of priority areas of scientific researches",  the project of the
Academy of Sciences of Ukraine, Code 6541230.


\end{document}